\documentclass[12pt]{article}
\usepackage{epsfig,cite}
\usepackage{amsmath}
\usepackage{amssymb}
\usepackage{color}
\usepackage{graphicx}
\usepackage{verbatim,mathrsfs,subfigure,feynmf}
\usepackage{url}
\include{epsf}
\newcount\timecount
\newcount\hours \newcount\minutes  \newcount\temp \newcount\pmhours
\hours = \time
\divide\hours by 60
\temp = \hours
\multiply\temp by 60
\minutes = \time
\advance\minutes by -\temp
\def\hour{\the\hours}
\def\minute{\ifnum\minutes<10 0\the\minutes
            \else\the\minutes\fi}
\def\clock{
\ifnum\hours=0 12:\minute\ AM
\else\ifnum\hours<12 \hour:\minute\ AM
      \else\ifnum\hours=12 12:\minute\ PM
            \else\ifnum\hours>12
                 \pmhours=\hours
                 \advance\pmhours by -12
                 \the\pmhours:\minute\ PM
                 \fi
            \fi
      \fi
\fi
}

\def\monthname{\relax\ifcase\month 0/\or January\or February\or
   March\or April\or May\or June\or July\or August\or September\or
   October\or November\or December\else\number\month/\fi}

\def\bold#1{\setbox0=\hbox{$#1$}%
     \kern-.025em\copy0\kern-\wd0
     \kern.05em\copy0\kern-\wd0
     \kern-.025em\raise.0433em\box0 }

\def\beq{\begin{equation}}
\def\eeq{\end{equation}}

\def\ga{\mathrel{\raise.3ex\hbox{$>$\kern-.75em\lower1ex\hbox{$\sim$}}}}
\def\la{\mathrel{\raise.3ex\hbox{$<$\kern-.75em\lower1ex\hbox{$\sim$}}}}
\def\gev{{\rm \, Ge\kern-0.125em V}}
\def\tev{{\rm \, Te\kern-0.125em V}}
\def\gyr{{\rm \, G\kern-0.125em yr}}

\def\gappeq{\mathrel{\rlap {\raise.5ex\hbox{$>$}}
{\lower.5ex\hbox{$\sim$}}}}
\def\lappeq{\mathrel{\rlap{\raise.5ex\hbox{$<$}}
{\lower.5ex\hbox{$\sim$}}}}
\def\Toprel#1\over#2{\mathrel{\mathop{#2}\limits^{#1}}}

\def\m12{m_{1\!/2}}

\def\bea{\begin{eqnarray}}
\def\eea{\end{eqnarray}}

\usepackage{lscape}

\def\beq{\begin{equation}}
\def\eeq{\end{equation}}

\def\mgut{M_{GUT}}

\begin{document}

\begin{titlepage}
\pagestyle{empty}
\rightline{KCL-PH-TH/2016-52, CERN-PH-TH/2016-185}
\rightline{KIAS--P16059, UMN--TH--3601/16, FTPI--MINN--16/23}
\vspace{2cm}
\begin{center}
{\bf {\Large The Super-GUT CMSSM Revisited} }
\end{center}

\vspace{0.5cm}
\begin{center}
{\bf John~Ellis}$^{1,2}$,
{\bf Jason L. Evans}$^3$,
{\bf Azar Mustafayev}$^{4}$, \\ \vspace{0.1cm}
{\bf Natsumi Nagata}$^{4}$
and {\bf Keith~A.~Olive}$^{4}$
\vskip 0.2in
{\small {\it
$^1${Theoretical Physics and Cosmology Group, Department of Physics, \\ King's College London, Strand,
London~WC2R 2LS, UK}\\
$^2${Theoretical Physics Department, CERN, CH-1211 Geneva 23, Switzerland}\\
$^3${School of Physics, KIAS, Seoul 130-722, Korea}\\
$^4${William I. Fine Theoretical Physics Institute, School of Physics and Astronomy,\\
University of Minnesota, Minneapolis, MN 55455,\,USA}}}\\

\vspace{1cm}
{\bf Abstract}
\end{center}
{\small
We revisit minimal supersymmetric SU(5) grand unification (GUT) models
 in which the soft
 supersymmetry-breaking parameters of the minimal supersymmetric
 Standard Model (MSSM) are universal at some input scale, $M_{
 in}$, above the supersymmetric gauge coupling unification scale,
 $M_{GUT}$. As in the constrained MSSM (CMSSM), we assume that the
 scalar masses and gaugino masses have common values, $m_0$ and
 $m_{1/2}$, respectively, at $M_{in}$, as do the trilinear soft
 supersymmetry-breaking parameters $A_0$. Going beyond previous studies
 of such a super-GUT CMSSM scenario, we explore the constraints imposed
 by the lower limit on the proton lifetime and the LHC measurement of the Higgs mass,
 $m_h$. We find regions of $m_0$, $m_{1/2}$, $A_0$ and the
 parameters of the SU(5) superpotential that are compatible with these
 and other phenomenological constraints such as the density of cold dark
 matter, which we assume to be provided by the lightest
 neutralino. Typically, these allowed regions appear for $m_0$ and
 $m_{1/2}$ in the multi-TeV region, for suitable values of
 the unknown SU(5) GUT-scale phases and superpotential couplings, and with the ratio of supersymmetric
 Higgs vacuum expectation values $\tan \beta \lesssim 6$.}


\vfill
\leftline{August 2016}
\end{titlepage}

\section{Introduction}
\label{sec:intro}

There have been many phenomenological studies of the minimal
supersymmetric (SUSY) extension of the Standard Model (MSSM) that assume some
degree of universality for the soft supersymmetry-breaking scalar and
gaugino masses, $m_0$ and $m_{1/2}$, and the trilinear soft
supersymmetry-breaking parameters $A_0$. Scenarios in which these
parameters are universal at the supersymmetric grand unification (GUT)
scale, $M_{GUT}$, called the constrained MSSM (CMSSM) \cite{funnel,
cmssm, efgo, cmssmwmap, eo6, ehow+}, have been particularly intensively
studied, usually assuming that the lightest supersymmetric particle
(LSP) is a neutralino, which is stable because of the conservation of
$R$-parity \cite{EHNOS}, and provides (all or some of) the cosmological cold
dark matter. These and other GUT-universal models are under strong
pressure from LHC data \cite{eo6, ehow+, mc75, mc8, mc9, ELOS, post-mh,
fp, eoz, elo, yal, Buchmueller:2015uqa, Ellis:2015rya}, in particular, the notable
absence of missing transverse energy signals at the LHC
\cite{ATLAS20, CMS20}, with the measurement of the Higgs mass \cite{lhch, 125}, $m_h$,
providing an additional important constraint.

Fewer studies have been performed for scenarios in which the soft
supersymmetry-breaking parameters are universal at some other scale
$M_{in} \ne M_{GUT}$, which might be either below the GUT scale
(so-called sub-GUT or GUT-less scenarios \cite{subGUT,ELOS,Ellis:2015rya}) or above the GUT
scale (so-called super-GUT scenarios \cite{superGUT,emo,dlmmo}). For
example, in our current state of confusion about the possible mechanism
of supersymmetry breaking, and specifically in the absence of a
convincing dynamical origin at $M_{GUT}$, one could well imagine that
the universality scale $M_{in}$ might lie closer to the Planck or string
scale: $M_{in} > M_{GUT}$.

When studying such super-GUT scenarios, there appear additional
ambiguities beyond those in the conventional CMSSM. What is $M_{in}$?
Which GUT model to study? What are its additional parameters? How much
additional freedom do they introduce? In parallel, once one commits to a
specific GUT model, one must also consider the constraint imposed by the
absence (so far) of proton decay \cite{Takhistov:2016eqm}. In order to
minimize the ambiguities and the number of additional GUT parameters,
we study here the minimal supersymmetric SU(5) GUT
\cite{Dimopoulos:1981zb}.

It is well known that the length of the proton lifetime
is a significant challenge for this model \cite{Goto:1998qg, mp}, and one of the
principal new ingredients in this paper, compared to previous studies of
super-GUT CMSSM models, is the incorporation of this constraint in our
exploration of the model parameter space. Another improvement on
previous super-GUT CMSSM studies is the incorporation of LHC constraints,
of which the measurement of the Higgs mass turns out to be the most
relevant.

We find regions of the soft supersymmetry-breaking parameters $m_0$,
$m_{1/2}$, $A_0$ and the unknown coefficients in the SU(5)
superpotential that are compatible with these and other phenomenological
constraints such as the density of cold dark matter. As usual, we assume
that this is provided by the LSP, which we assume to be the lightest
neutralino. The Higgs mass and proton lifetime constraints both favor
$m_0$ and $m_{1/2}$ in the multi-TeV region, and proton stability
favours a value $\lesssim 6$ for the ratio of supersymmetric Higgs
vacuum expectation values (VEVs), $\tan \beta$. The cosmological
constraint on the cold dark matter density typically favors narrow
strips of parameter space where coannihilation with the lighter stop
brings the LSP density into the cosmological range. All these
constraints can be reconciled for suitable values of the unknown SU(5)
superpotential couplings.

The layout of this paper is as follows. In Section~\ref{sec:supergut}
we review our set-up of the super-GUT CMSSM, with particular attention
to the model parameters and the matching to the relevant parameters
below the GUT scale. Section~\ref{pdecay} then reviews our treatment
of proton decay, paying particular attention to the potential implications
of unknown GUT-scale phases. Our results are presented and explained in
Section~\ref{sec:results}, and Section~\ref{sec:discussion} then
summarizes our conclusions. An Appendix reviews details of our
nucleon decay calculations.

\section{Super-GUT CMSSM Models}
\label{sec:supergut}

\subsection{Minimal SUSY SU(5)}

We first review briefly the minimal supersymmetric SU(5) GUT~\cite{Dimopoulos:1981zb},
specifying our notation. This model is the
simplest supersymmetric extension of the original SU(5) GUT model due to
Georgi and Glashow \cite{Georgi:1974sy}. In this model, the right-handed
down-type quark and left-handed lepton chiral superfields, $\overline{D}_i$
and ${L}_i$, respectively, reside in $\bf{\overline{5}}$
representations, ${\Phi}_i$, while the left-handed quark doublet, right-handed
up-type quark, and right-handed charged-lepton chiral superfields,
${Q}_i$, $\overline{U}_i$, and $\overline{E}_i$, respectively, are in
$\bf{10}$ representations, ${\Psi}_i$, where the index $i = 1,2,3$ denotes the
generations. The MSSM Higgs chiral superfields
$H_u$ and $H_d$ are embedded into ${\bf 5}$ and $\overline{\bf 5}$
representations, $H$ and $\overline{H}$, respectively, where they are accompanied by
the ${\bf 3}$ and $\overline{\bf 3}$ coloured Higgs superfields $H_C$ and $\overline{H}_C$, respectively.

The SU(5) GUT gauge symmetry is assumed to be spontaneously broken down to the Standard
Model (SM) gauge group by the vacuum expectation value (vev) of a ${\bf 24}$ chiral superfield,
$\Sigma \equiv \sqrt{2}\Sigma^A T^A$, where the $T^A$ ($A=1, \dots, 24$) are the
generators of SU(5) normalized so that ${\rm Tr}(T^A T^B) =
\delta_{AB}/2$. The renormalizable superpotential for this model is then
given by
\begin{align}
 W_5 &=  \mu_\Sigma {\rm Tr}\Sigma^2 + \frac{1}{6} \lambda^\prime {\rm
 Tr} \Sigma^3 + \mu_H \overline{H} H + \lambda \overline{H} \Sigma H
\nonumber \\
&+ \left(h_{\bf 10}\right)_{ij} \epsilon_{\alpha\beta\gamma\delta\zeta}
 \Psi_i^{\alpha\beta} \Psi^{\gamma\delta}_j H^\zeta +
 \left(h_{\overline{\bf 5}}\right)_{ij} \Psi_i^{\alpha\beta} \Phi_{j \alpha}
 \overline{H}_\beta ~,
\label{W5}
\end{align}
where Greek sub- and superscripts denote SU(5) indices, and $\epsilon$ is the totally
antisymmetric tensor with $\epsilon_{12345}=1$.

The adjoint Higgs $\Sigma$ is assumed to have a vev of the form
\begin{equation}
 \langle \Sigma \rangle = V \cdot {\rm diag} \left(2,2,2,-3,-3\right) ~,
\end{equation}
where $V\equiv 4 \mu_\Sigma/\lambda^\prime$.
In this case, the GUT gauge bosons acquire masses $M_X = 5 g_5 V$,
where $g_5$ is the SU(5) gauge coupling. In order to realize the
doublet-triplet mass splitting in $H$ and $\overline{H}$, we need to
impose the fine-tuning condition $\mu_H -3\lambda V \ll V$, which we
discuss in Section~\ref{sec:matchingcond}. In this case, the masses of the
color and weak adjoint components of $\Sigma$ are equal to $M_\Sigma =
5\lambda^\prime V/2$, while the singlet component of $\Sigma$ acquires a
mass $M_{\Sigma_{24}} = \lambda^\prime V/2$. The color-triplet Higgs
states have masses $M_{H_C} = 5\lambda V$.

\subsection{Planck-scale suppressed higher-dimensional operators}
\label{eq:highdimops}

In supersymmetric GUTs, gauge-coupling unification predicts that the unification scale
is ${\cal O}(10^{16})$~GeV. Since the unification scale is
fairly close to the reduced Planck mass $M_P = 2.4\times 10^{18}$~GeV,
interactions of gravitational strength may give rise to sizable effects. We
accommodate these effects by considering higher-dimensional effective
operators suppressed by powers of $M_P$.

We may expect that such effective
operators play significant roles in the minimal SUSY SU(5) GUT. For example, in minimal SU(5) GUTs
fthe down-type Yukawa couplings are predicted to be equal to the
corresponding lepton Yukawa couplings at the GUT scale, since they both
originate from $h_{\overline{\bf 5}}$. Nevertheless, in most of the
parameter space we consider, this Yukawa unification is imperfect. For
the third generation, the deviation is typically at the ${\cal O}(10)$\%
level. For the first two generations, on the other hand, there are ${\cal
O}(1)$ differences. These less successful predictions can be rectified
if one considers the following dimension-five effective operators
that are suppressed by the Planck scale~\cite{nro, Bajc:2002pg}:
\begin{equation}
 W_{\rm eff}^{\Delta h} = \frac{c^{\Delta h}_{ij}}{M_P} \Phi_{i \alpha}
  \Sigma^\alpha_{~\beta} \Psi^{\beta\gamma} \overline{H}_\gamma ~.
\label{eq:weffdelh}
\end{equation}
These operators induce non-universal contributions to the effective Yukawa couplings that are ${\cal O}(V/M_P)$ after the
adjoint Higgs acquires a VEV~\footnote{There is another class of
dimension-five operators of the form $\Psi_i^{\alpha\beta} \Phi_{j\alpha}
\Sigma^\gamma_{~\beta} \overline{H}_\gamma$. However, they do not spoil
Yukawa unification, but only modify the overall sizes of the down-type
quark and charged-lepton Yukawa couplings by ${\cal O}(V/M_P)$.},
which is sufficient to account for the observed deviations~\footnote{One
may also use higher-dimensional Higgs representations to explain the
observed differences between down-type and lepton Yukawa
couplings \cite{GM}. However, in this paper we focus on the minimal SU(5) GUT,
and do not consider this alternative.}.

There are several other dimension-five operators that one may consider. Among them is
\begin{equation}
 W_{\rm eff}^{\Delta g} = \frac{c}{M_P} {\rm Tr}\left[
\Sigma {\cal W} {\cal W}
\right] ~,
\label{eq:SigmaWW}
\end{equation}
where ${\cal W}\equiv T^A {\cal W}^A$ denotes the
superfields corresponding to the field strengths of the SU(5) gauge vector bosons
${\cal V} \equiv {\cal V}^A T^A$. The term (\ref{eq:SigmaWW}) can have a significant effect, since it
changes the matching conditions of the gauge coupling constants
after $\Sigma$ develops a VEV \cite{Ellis:1985jn, Hill:1983xh,
Tobe:2003yj}. This operator also modifies the matching conditions for
gaugino masses, thereby modifying gaugino mass unification
\cite{Ellis:1985jn, Tobe:2003yj, Anderson:1996bg}. We discuss these
effects in detail in Section~\ref{sec:matchingcond}.

We may also have terms of the form~\cite{Bajc:2002pg}
\begin{equation}
 W_{\rm eff}^\Sigma = \frac{a}{M_P} \left({\rm Tr} \Sigma^2\right)^2
+\frac{b}{M_P} {\rm Tr} \Sigma^4 ~.
\end{equation}
These operators can split the masses of the color and SU(2)$_L$ adjoint
components in $\Sigma$, $M_{\Sigma_8}$ and $M_{\Sigma_3}$ by ${\cal
O}(V^2/M_P)$. This mass difference induces threshold corrections to gauge
coupling constants of  $\sim
\ln(M_{\Sigma_3}/M_{\Sigma_8})/(16\pi^2)$. This effect is negligible
for $\lambda^\prime \gg (a,b)V/M_P$ but could be significant for very small $\lambda^\prime$.
However, in order to simplify our analysis, we neglect the effects of these operators in this
paper.

\subsection{Soft supersymmetry-breaking mass parameters}

The soft supersymmetry-breaking terms in the minimal supersymmetric SU(5) GUT are
\begin{align}
 {\cal L}_{\rm soft} = &- \left(m_{\bf 10}^2\right)_{ij}
 \widetilde{\psi}_i^* \widetilde{\psi}_j
- \left(m_{\overline{\bf 5}}^2\right)_{ij} \widetilde{\phi}^*_i
 \widetilde{\phi}_j
- m_H^2 |H|^2 -m_{\overline{H}}^2 |\overline{H}|^2 - m_\Sigma^2 {\rm Tr}
\left(\Sigma^\dagger \Sigma\right)
\nonumber \\
&-\biggl[
\frac{1}{2}M_5 \widetilde{\lambda}^{A} \widetilde{\lambda}^A
+ A_{\bf 10} \left(h_{\bf 10}\right)_{ij}
 \epsilon_{\alpha\beta\gamma\delta\zeta} \widetilde{\psi}_i^{\alpha\beta}
 \widetilde{\psi}^{\gamma\delta}_j H^\zeta
+ A_{\overline{\bf 5}}\left(h_{\overline{\bf 5}}\right)_{ij}
 \widetilde{\psi}_i^{\alpha\beta} \widetilde{\phi}_{j \alpha}  \overline{H}_\beta
\nonumber \\
&+ B_\Sigma \mu_\Sigma {\rm Tr} \Sigma^2 +\frac{1}{6} A_{\lambda^\prime
 } \lambda^\prime  {\rm Tr} \Sigma^3 +B_H \mu_H \overline{H} H+
 A_\lambda \lambda \overline{H} \Sigma H +{\rm h.c.}
 \biggr]~,
\end{align}
where $\widetilde{\psi}_i$ and $\widetilde{\phi}_i$ are the scalar
components of $\Psi_i$ and $\Phi_i$, respectively,
the $\widetilde{\lambda}^A$ are the SU(5) gauginos, and we
use the same symbols for the scalar components of the Higgs fields as for the
corresponding superfields.

In the super-GUT CMSSM model, we impose the following universality
conditions for the soft-mass parameters at a soft supersymmetry-breaking mass input scale
$M_{in} > M_{\rm GUT}$:
\begin{align}
 \left(m_{\bf 10}^2\right)_{ij} =
\left(m_{\overline{\bf 5}}^2\right)_{ij}
&\equiv m_0^2 \, \delta_{ij} ~,
\nonumber \\[3pt]
m_H = m_{\overline{H}} = m_\Sigma &\equiv m_0 ~,
\nonumber \\[3pt]
A_{\bf 10} = A_{\overline{\bf 5}} = A_\lambda = A_{\lambda^\prime}
&\equiv A_0 ~,
\nonumber \\[3pt]
 M_5 &\equiv m_{1/2} ~.
\label{eq:inputcond}
\end{align}
The bilinear soft SUSY-breaking therms $B_\Sigma$ and $B_H$ are determined from the other
parameters, as we shall see in the following. Note that, if we set $M_{in} =
M_{GUT}$, the above conditions are equivalent to those in the CMSSM.

These parameters are evolved down to $M_{GUT}$ using the
renormalization-group equations (RGEs) of the minimal supersymmetric SU(5) GUT,
which can be found in~\cite{pp,Baer:2000gf,emo}, with appropriate changes
of notation. During the evolution, the GUT parameters in Eq.~\eqref{W5}
affect the running of the soft supersymmetry-breaking parameters, which results
in non-universality in the soft parameters at $M_{GUT}$. In particular,
the $\lambda$ coupling enters into the RGEs for the soft masses of the
${\bf 5}$ and $\overline{\bf 5}$ Higgs fields, and can have
significant effects on their evolution. These effects become
particularly important in the vicinity of the focus-point region at
large $m_0$, since it is very close to the boundary of consistent
electroweak symmetry breaking (EWSB). In addition, $\lambda$ contributes
to the running of the Yukawa couplings and the corresponding $A$-terms.
On the other hand,
$\lambda^\prime$ affects directly only the running of $\lambda$,
$m_\Sigma$, and $A_\lambda$ (besides $\lambda^\prime$ and
$A_{\lambda^\prime}$), and thus can affect the MSSM soft mass parameters
only at higher-loop level. Both of $\lambda$ and
$\lambda^\prime $ contribute to the RGEs of the soft masses of
matter multiplets only at higher-loop level, and thus their effects on
these parameters are rather small. Thus, the low-energy phenomenology is rather
insensitive to the value of $\lambda^\prime$.  The $\mu$ parameters $\mu_\Sigma$ and
$\mu_H$, as well as the corresponding bilinear parameters $B_\Sigma$ and
$B_H$, do not enter into RGEs of the rest of the parameters, and thus
their values give no effects on the running of the parameters in
Eq.~\eqref{eq:inputcond}. We note in passing that, if we set $M_{in} =
M_{GUT}$, we obtain the CMSSM and there is no effect from the running
above the GUT scale on the low-energy spectrum \footnote{However, we find
that the GUT-scale matching condition on the $B$ parameter gives a
constraint on the model parameter space even though $M_{in} = M_{GUT}$,
as we see below. }.

\subsection{GUT-scale matching conditions}
\label{sec:matchingcond}

At the unification scale $M_{GUT}$, the SU(5) GUT parameters are matched
onto the MSSM parameters. In this Section, we summarize these matching conditions and
discuss the constraints on the parameters from the low-energy
observables.

The matching conditions for the Standard Model gauge couplings at one-loop level in
the $\overline{\rm DR}$ scheme are given by
\begin{align}
 \frac{1}{g_1^2(Q)}&=\frac{1}{g_5^2(Q)}+\frac{1}{8\pi^2}\biggl[
\frac{2}{5}
\ln \frac{Q}{M_{H_C}}-10\ln\frac{Q}{M_X}
\biggr]+\frac{8cV}{M_P} (-1)
~, \label{eq:matchg1} \\
 \frac{1}{g_2^2(Q)}&=\frac{1}{g_5^2(Q)}+\frac{1}{8\pi^2}\biggl[
2\ln \frac{Q}{M_\Sigma}-6\ln\frac{Q}{M_X}
\biggr]+\frac{8cV}{M_P} (-3)
~, \\
 \frac{1}{g_3^2(Q)}&=\frac{1}{g_5^2(Q)}+\frac{1}{8\pi^2}\biggl[
\ln \frac{Q}{M_{H_C}}+3\ln \frac{Q}{M_\Sigma}-4\ln\frac{Q}{M_X}
\biggr]+\frac{8cV}{M_P} (2)~,
\end{align}
where $g_1$, $g_2$, and $g_3$ are the U(1), SU(2), and SU(3) gauge
couplings, respectively, and $Q$ is a renormalization scale taken in our analysis to be
the unification scale: $Q = M_{GUT}$. The last terms in
these equations represent the contribution of the dimension-five
operator \eqref{eq:SigmaWW}. Since $V/M_P \simeq 10^{-2}$, these terms
can be comparable to the one-loop threshold corrections, and thus should
be taken into account when discussing gauge-coupling unification~\cite{Tobe:2003yj}.
From these equations, we have
\begin{align}
 \frac{3}{g_2^2(Q)} - \frac{2}{g_3^2(Q)} -\frac{1}{g_1^2(Q)}
&=-\frac{3}{10\pi^2} \ln \left(\frac{Q}{M_{H_C}}\right)
-\frac{96cV}{M_P}
~,\label{eq:matchmhc} \\[3pt]
 \frac{5}{g_1^2(Q)} -\frac{3}{g_2^2(Q)} -\frac{2}{g_3^2(Q)}
&= -\frac{3}{2\pi^2}\ln\left(\frac{Q^3}{M_X^2 M_\Sigma}\right) ~,
\label{eq:matchmgut}
\\[3pt]
 \frac{5}{g_1^2(Q)} +\frac{3}{g_2^2(Q)} -\frac{2}{g_3^2(Q)}&= -\frac{15}{2\pi^2} \ln\left(\frac{Q}{M_X}\right) + \frac{6}{g_5^2(Q)} -\frac{144cV}{M_P} ~,\label{eq:matchg5}
\end{align}
We note that there is no contribution to (\ref{eq:matchmgut}) from the
dimension-five operator~\footnote{This feature can be understood as
follows. The contributions of the color-triplet Higgs multiplets to the
gauge coupling beta functions are given by $(b_1^{H_C}, b_2^{H_C},
b_3^{H_C})  =(2/5, 0, 1)$. In this notation, the matching conditions may be rewritten as
\begin{equation}
 \frac{1}{g_i^2 (Q)} = \frac{1}{g_5^2(Q)} +\frac{1}{8\pi^2}
\left[b_i^{H_C} \ln\left(\frac{Q}{M_{H_C}}\right)+ \dots\right]
+\frac{8cV}{M_{P}}\left(-3 + 5b_i^{H_C}\right) ~.
\end{equation}
Since $5b_1^{H_C} -3b_{2}^{H_C}-2b^{H_C}_3 = 0$ and $5-3-2=0$,
neither $\ln (M_{H_C})$ nor $V/M_P$ appears in (\ref{eq:matchmgut}).}.
By running the gauge couplings up from their low-energy values, we can
determine the combination $M_X^2 M_\Sigma$ via
(\ref{eq:matchmgut})~\cite{Hisano:1992mh, Hisano:1992jj,
Hisano:2013cqa}. Notice that without the dimension-five operator ($c=0$),
$M_{H_C}$ is also determined from the values of the gauge couplings at
the GUT scale via Eq.~\eqref{eq:matchmhc}. The contribution of this
operator relaxes this constraint, and allows us to regard $M_{H_C}$ as a
free parameter. The last matching condition, Eq. (\ref{eq:matchg5}), will be used to determine $g_5$ and $M_{H_C}$ as will be discussed below.

For the Yukawa couplings, we use the tree-level matching
conditions. However, we note here that there is an ambiguity in the determination
of the GUT Yukawa couplings. As we mentioned in
Section~\ref{eq:highdimops}, Yukawa unification in the MSSM is imperfect in
most of the parameter space. Although this is cured by the
higher-dimensional operators in (\ref{eq:weffdelh}), they introduce
additional contributions to the matching conditions for the Yukawa
couplings. With this in mind, in this paper, we use
\begin{equation}
 h_{{\bf 10}, 3} = \frac{1}{4}f_{u_3}~, ~~~~~~
 h_{\overline{\bf 5}, 3} = \frac{f_{d_3} + f_{e_3}}{\sqrt{2}} ~,
\end{equation}
for the third-generation Yukawa couplings,
where $h_{{\bf 10}, i}$, $h_{\overline{\bf 5}, i}$, $f_{u_i}$,
$f_{d_i}$, and $f_{e_i}$ are eigenvalues of $h_{\bf 10}$,
$h_{\overline{\bf 5}}$, the MSSM up-type Yukawa couplings, the MSSM
down-type Yukawa couplings, and the MSSM lepton Yukawa couplings,
respectively. This condition is the same as that used in
Ref.~\cite{emo}. For the first- and second-generation Yukawa couplings,
on the other hand, we use
\begin{equation}
 h_{{\bf 10}, i} = \frac{1}{4} f_{u_i} ~, ~~~~~~
 h_{\overline{\bf 5}, i} = \sqrt{2} f_{d_i} ~.
\end{equation}
We chose the down-type Yukawa
couplings for the $h_{\overline{\bf 5}}$ matching condition, rather than
the lepton Yukawa couplings, since it results in longer proton decay lifetimes
and thus gives a conservative bounds on the model parameter space
\cite{Ellis:2015rya, evno}.

Next we obtain the matching conditions for the soft supersymmetry-breaking terms. To
this end, we first note that in the presence of soft supersymmetry-breaking terms
the VEV of $\Sigma$ deviates from $V$ by ${\cal O}(M_{\rm SUSY})$, where
$M_{\rm SUSY}$ denotes the supersymmetry-breaking scale \cite{Hall:1983iz}. In addition, $\langle
\Sigma \rangle$ develops a non-vanishing $F$-term. We find that
\begin{equation}
 \langle \Sigma \rangle =
\left[
V + \frac{V(A_{\lambda^\prime} - B_\Sigma)}{2\mu_\Sigma}
+ F_\Sigma\, \theta^2
\right]\cdot {\rm diag}(2,2,2,-3,-3) ~,
\end{equation}
where
\begin{equation}
 F_\Sigma = V(A_{\lambda^\prime} - B_\Sigma)
+\frac{V}{2\mu_\Sigma}
\left[
B_\Sigma(A_{\lambda^\prime} -B_\Sigma)
-m_\Sigma^2
\right] + {\cal O}(M_{\rm SUSY}^3/M_{GUT})~.
\end{equation}
Using this result, we obtain the following matching conditions for the gaugino
masses \cite{Tobe:2003yj, Hisano:1993zu}:
\begin{align}
 M_1 &= \frac{g_1^2}{g_5^2} M_5
-\frac{g_1^2}{16\pi^2}\left[10 M_5 +10(A_{\lambda^\prime} -B_\Sigma)
 +\frac{2}{5}B_H\right]
+\frac{4cg_1^2V(A_{\lambda^\prime} -B_\Sigma)}{M_P} ~,
\label{eq:m1match}
\\[3pt]
M_2 &= \frac{g_2^2}{g_5^2} M_5
-\frac{g_2^2}{16\pi^2}\left[6 M_5 +6A_{\lambda^\prime} -4B_\Sigma
 \right]
+\frac{12cg_2^2V(A_{\lambda^\prime} -B_\Sigma)}{M_P} ~,
\label{eq:m2match}
\\[3pt]
M_3 &= \frac{g_3^2}{g_5^2} M_5
-\frac{g_3^2}{16\pi^2}\left[4 M_5 +4A_{\lambda^\prime} -B_\Sigma
+B_H \right]
-\frac{8cg_3^2V(A_{\lambda^\prime} -B_\Sigma)}{M_P}
~.
\label{eq:m3match}
\end{align}
We again find that the contribution of the dimension-five operator
can be comparable to that of the one-loop threshold corrections.

The soft masses of the MSSM matter fields, as well as the $A$-terms of
the third-generation sfermions, are given by 
\begin{align}
 m^2_{Q} = m_{U}^2 = m^2_{E} = m^2_{{\bf 10}} ~,&
~~~~~~ m_{D}^2 = m_{L}^2 = m_{\overline{\bf 5}}^2 ~, \nonumber \\
 m_{H_u}^2 = m_H^2 ~,& ~~~~~~ m_{H_d}^2 = m_{\overline{H}}^2 ~,
\nonumber \\
 A_t = A_{\bf 10} ~,& ~~~~~~
 A_b = A_\tau = A_{\overline{\bf 5}} ~.
\end{align}

Finally, for the $\mu$ and $B$ terms we have \cite{Borzumati:2009hu}
\begin{align}
 \mu &= \mu_H - 3 \lambda V\left[
1+ \frac{A_{\lambda^\prime} -B_\Sigma}{2 \mu_\Sigma}
\right] ~,
\label{eq:matchingmu}
 \\[3pt]
 B &= B_H + \frac{3\lambda V \Delta}{\mu}
+ \frac{6 \lambda}{\lambda^\prime \mu} \left[
(A_{\lambda^\prime} -B_\Sigma) (2 B_\Sigma -A_{\lambda^\prime}
 +\Delta) -m_\Sigma^2
\right]~,
\label{eq:matchingb}
\end{align}
with
\begin{equation}
 \Delta \equiv A_{\lambda^\prime} - B_\Sigma - A_\lambda +B_H ~.
\label{eq:deltadef}
\end{equation}
These equations display the amount of fine-tuning required to obtain
values of $\mu$ and $B$ that are ${\cal O}(M_{\rm SUSY})$. Equation
\eqref{eq:matchingmu} shows that we need to tune $|\mu_H -3\lambda V|$
to be ${\cal O}(M_{\rm SUSY})$. On the other hand,
Eq.~\eqref{eq:matchingb} indicates that $V\Delta/\mu$ should be ${\cal
O}(M_{\rm SUSY})$, which requires $|\Delta| \leq {\cal O}(M_{\rm
SUSY}^2/M_{GUT})$. Therefore, we can neglect $\Delta$ in the following
calculations. Notice that the condition $\Delta = 0$ is stable against radiative corrections
as shown in Ref.~\cite{Kawamura:1994ys}.

The $\mu$ and $B$ parameters are determined by using the electroweak
vacuum conditions:
\begin{align}
 \mu^2 &= \frac{m_{1}^2 -m_{2}^2 \tan^2\beta + \frac{1}{2} m_Z^2
 (1-\tan^2\beta ) +\Delta_\mu^{(1)}}{\tan^2 \beta -1 +\Delta_\mu^{(2)}}
 , \\
 B\mu &= -\frac{1}{2}(m_1^2 + m_2^2 +2 \mu^2) \sin 2\beta +\Delta_B ~,
\end{align}
where $\Delta_B$ and $\Delta_\mu^{(1,2)}$ denote loop corrections
\cite{Barger:1993gh}.

We can determine the $B$ parameters in minimal SU(5)
by solving the conditions \eqref{eq:matchingb} and $\Delta = 0$
\footnote{We need to determine the $B$ parameters in order to obtain
the MSSM gaugino masses via Eqs.~(\ref{eq:m1match}--\ref{eq:m3match}).}.
However, we
find that there is an additional condition that must be satisfied in order for
these equations to be solvable. When eliminating $B_H$ from
Eq.~\eqref{eq:matchingb} using $\Delta = 0$, we obtain an
equation that is quadratic in $B_\Sigma$. This equation has a
real solution only if
\begin{equation}
 A_{\lambda^\prime}^2 -\frac{\lambda^\prime \mu}{3\lambda}
\left(A_{\lambda^\prime} -4 A_\lambda + 4B\right)
+\left(\frac{\lambda^\prime \mu}{6\lambda}\right)^2
\geq 8 m_{\Sigma}^2 ~.
\label{alimit}
\end{equation}
This condition gives a non-trivial constraint on the input parameters,
especially on the trilinear coupling $A_0$. In particular, for
$\lambda^\prime \ll \lambda$, this constraint leads to
$A_{\lambda^\prime}^2 \simeq A_0^2 \geq 8 m_{\Sigma}^2 \simeq 8m_0^2$.

When we compute the proton lifetime, we need to evaluate the
color-triplet Higgs mass $M_{H_C}$. This can be done by using
Eqs.~\eqref{eq:matchmhc}, \eqref{eq:matchmgut}, and \eqref{eq:matchg5} together with
\begin{align}
 M_{H_C} &= 5\lambda V ~,\label{eq:MHCV} \\
 M_\Sigma &= \frac{5}{2} \lambda^\prime V ~, \\
 M_X &= 5 g_5 V ~.\label{eq:MXV}
\end{align}
From these equations, we obtain
\begin{equation}
 M_{H_C} = \lambda \left(\frac{2}{\lambda^\prime
		    g_5^2}\right)^{\frac{1}{3}}
\left(M_X^2 M_\Sigma\right)^{\frac{1}{3}} ~.
\label{eq:mhc}
\end{equation}
We can then determine $M_X^2 M_\Sigma$ using Eq.~\eqref{eq:matchmgut}. Eq. (\ref{eq:matchg5}) can be reduced to an equation with undetermined parameters $g_5$ and $M_{H_C}$ using Eq.  (\ref{eq:MHCV}) and (\ref{eq:MXV}). Then once $\lambda$ and $\lambda^\prime$  are chosen, this equation plus Eq.~\eqref{eq:mhc} can be used
to determine $M_{H_C}$ and $g_5$. However, since $g_5$ is only logarithmical dependent on $M_{H_C}$, it will remain fairly constant for a broad range of $M_{H_C}$. As mentioned above, if we do not include the
contribution of the dimension-five operator, Eq.~\eqref{eq:matchmhc}
fixes $M_{H_C}$. In this case, $\lambda$ and $\lambda^\prime$ are restricted via
Eq.~\eqref{eq:mhc}, and thus we cannot regard both of them as free
parameters. The last term in Eq.~\eqref{eq:matchmhc} can relax this
restriction, and enables us to take $\lambda$ and $\lambda^\prime$ as
input parameters. In this case, $M_{H_C}$ is given by
Eq.~\eqref{eq:mhc}, and Eq.~\eqref{eq:matchmhc} determines the parameter
$c$. In the following analysis, we check that the coefficient $c$
has reasonable values, {\it i.e.}, $|c| < {\cal O} (1)$.

Using the above results, we see how the super-GUT CMSSM model is
specified by the following set of input parameters:
\begin{equation}
 m_0,\ m_{1/2},\ A_0,\ M_{in},\ \lambda,\ \lambda',\ \tan \beta,\ {\rm
  sign}(\mu) \, ,
\end{equation}
where the trilinear superpotential Higgs couplings, $ \lambda,\ \lambda'$,
are specified at $Q=\mgut$.

\section{Proton Decay and GUT-Scale Phases}
\label{pdecay}

As is well known, in the minimal supersymmetric SU(5) GUT with weak-scale
supersymmetry breaking, the dominant decay channel
of proton is the $p\to K^+ \overline{\nu}$ mode \cite{Sakai:1981pk},
which is induced by the exchange of the color-triplet Higgs
multiplets, and the model is severely restricted by the proton decay
bound \cite{Goto:1998qg, mp}. The exchange of the GUT-scale gauge bosons can also induce
proton decay, but this contribution is usually subdominant because of the large
GUT scale in supersymmetric GUTs. The strong constraint from the
$p\to K^+ \overline{\nu}$ decay may, however, be evaded if the masses of
supersymmetric particles are well above the electroweak scale
\cite{Hisano:2013exa, McKeen:2013dma, Nagata:2013sba, evno,
Ellis:2015rya}. In addition, it turns out that the $p\to K^+
\overline{\nu}$ decay mode depends sensitively on the extra phases in the GUT
Yukawa couplings \cite{Ellis:1979hy}, which can suppress the
proton decay rate, as we discuss in this Section.
For more details of the proton decay calculation, see
Refs.~\cite{Hisano:2013exa, Nagata:2013sba, evno, Ellis:2015rya} and
the Appendix.

In supersymmetric models, the largest contribution to the decay rate of
the proton is determined by the dimension-five effective operators
generated by integrating out the colored Higgs multiplets
\cite{Sakai:1981pk},
\begin{equation}
{\cal L}_5^{\rm eff}= C^{ijkl}_{5L}{\cal O}^{5L}_{ijkl}
+C^{ijkl}_{5R}{\cal O}^{5R}_{ijkl}
+{\rm h.c.}~,
\label{eq:efflaggut}
\end{equation}
with ${\cal O}^{5L}_{ijkl}$ and ${\cal
O}^{5R}_{ijkl}$ defined by
\begin{align}
 {\cal O}^{5L}_{ijkl}&\equiv\int d^2\theta~ \frac{1}{2}\epsilon_{abc}
(Q^a_i\cdot Q^b_j)(Q_k^c\cdot L_l)~,\nonumber \\
{\cal O}^{5R}_{ijkl}&\equiv\int d^2\theta~
\epsilon^{abc}\overline{u}_{ia}\overline{e}_j\overline{u}_{kb}
\overline{d}_{lc}~,
\end{align}
where $i,j,k,l$ are generation indices, $a,b,c$ are SU(3)$_C$ color
indices, and $\epsilon_{abc}$ is the totally antisymmetric three-index tensor.
The Wilson coefficients are given by
\begin{align}
 C^{ijkl}_{5L}(M_{GUT})&
=\frac{2\sqrt{2}}{M_{H_C}}h_{{\bf 10}, i}
e^{i\phi_i}\delta^{ij}V^*_{kl}h_{\overline{\bf 5},l}~,\nonumber \\
C^{ijkl}_{5R}(M_{GUT})
&=\frac{2\sqrt{2}}{M_{H_C}}h_{{\bf 10}, i}V_{ij}V^*_{kl}h_{\overline{\bf 5},l}
e^{-i\phi_k}
~,
\label{eq:wilson5}
\end{align}
where $V_{ij}$ are the familiar CKM matrix elements, and the $\phi_i$ ($i =1,2,3$) are
the new CP- violating phases in the GUT Yukawa couplings. These are subject to the
constraint $\phi_1 + \phi_2 + \phi_3 = 0$, so there are two independent
degrees of freedom for these new CP-violating phases \cite{Ellis:1979hy}~\footnote{The
number of extra degrees of freedom in the GUT Yukawa couplings can be counted as
follows. Since $h_{\bf 10}$ is a $3\times 3$ symmetric complex matrix,
it has 12 real degrees of freedom, while $h_{\overline{\bf 5}}$ has
18. Field redefinitions of $\Psi_i$ and $\Phi_i$ span the ${\rm U}(3) \otimes
{\rm U}(3)$ transformation group, and thus 18 parameters are
unphysical. Hence, we have 12 physical parameters. Among them, 6 are specified by
quark masses, while 4 are for the CKM matrix elements. The remaining 2
are the extra CP phases, which we take to be $\phi_2$ and $\phi_3$.}.
We take $\phi_2$ and $\phi_3$ as free input parameters in the following
discussion. The coefficients in Eq.~\eqref{eq:wilson5} are then run
to the SUSY scale using the RGEs. At the SUSY scale, the sfermions associated with these Wilson
coefficients are integrated out through a loop containing either a wino
mass insertion or a Higgsino mass insertion, which are proportional to
$C_{5L}$ and $C_{5R}$, respectively. The wino contribution to the decay
amplitude for the $p\to K^+ \overline{\nu}_i$ mode is given by the sum
of the Wilson coefficients $C_{LL} (usd\nu_i)$ and $C_{LL} (uds\nu_i)$
multiplied by the corresponding matrix elements (see
Eq.~\eqref{eq:amplitudes}). These coefficients are approximated by
\begin{align}
 C_{LL} (usd\nu_i) &= C_{LL} (uds\nu_i) \nonumber \\
&\simeq
\frac{2\alpha_2^2}{\sin 2\beta}\frac{m_t m_{d_i} M_2}{ m_W^2M_{H_C}M_{\rm
 SUSY}^2} V_{ui}^*V_{td}V_{ts}e^{i\phi_3}\left(1 +
 e^{i(\phi_2-\phi_3)}\frac{m_c V_{cd}V_{cs}}{m_tV_{td}V_{ts}}\right) ~,
\label{eq:cllaprox}
\end{align}
where $m_c$, $m_t$, $m_W$, and $m_{d_i}$ are the masses of the charm quark,
top quark, $W$ boson, and down-type quarks, respectively, and $\alpha_2 = g_2^2/4\pi$.
Since the ratio of Yukawa couplings and CKM matrix elements in the
parenthesis in Eq.~\eqref{eq:cllaprox} is ${\cal O}(1)$, this Wilson coefficient may be suppressed
for certain ranges of the phases.
On the other hand, the Higgsino exchange process contributes only to the
$p\to K^+ \overline{\nu}_\tau$ mode, and gives no contribution to
the $p\to K^+ \overline{\nu}_{e, \mu}$ modes. The relevant Wilson
coefficients for the $p\to K^+ \overline{\nu}_\tau$ mode are $C_{LL}
(usd\nu_\tau)$ and $C_{LL}
(uds\nu_\tau)$ in Eq.~\eqref{eq:cllaprox}, as well as $C_{RL}
(usd\nu_\tau)$ and $C_{RL} (uds \nu_\tau)$, which are approximately given by
\begin{align}
 C_{RL} (usd\nu_\tau) &\simeq -
 \frac{\alpha_2^2}{\sin^2 2\beta}\frac{m_t^2 m_s m_\tau \mu}{ m_W^4 M_{H_C}M_{\rm SUSY}^2}V_{tb}^*
 V_{us} V_{td}e^{-i(\phi_2+\phi_3)} ~, \nonumber \\
 C_{RL} (uds\nu_\tau) &\simeq -
 \frac{\alpha_2^2}{\sin^2 2\beta}\frac{m_t^2 m_d m_\tau \mu}{m_W^4 M_{H_C} M_{\rm SUSY}^2}V_{tb}^*
 V_{ud} V_{ts}e^{-i(\phi_2+\phi_3)} ~,
\label{eq:crlaprox}
\end{align}
where $m_d$, $m_s$, and $m_\tau$ are the masses of down quark, strange
quark, and tau lepton, respectively.
Contrary to the coefficients in Eq.~\eqref{eq:cllaprox}, the absolute
values of these coefficients do not change when the phases vary.

Equations \eqref{eq:cllaprox} and \eqref{eq:crlaprox} show that the
proton decay rate receives a $\tan \beta$ enhancement as well as
a suppression by the sfermion mass scale $M_{\rm SUSY}$. To evade
the proton decay bound, therefore, a small $\tan \beta$ and a high
supersymmetry-breaking scale are favored as shown in the subsequent
section. In addition, we note that the proton decay rate decreases as
$M_{H_C}$ is taken to be large. From Eq.~\eqref{eq:mhc}, we find
$M_{H_C} \propto \lambda/ (\lambda^\prime)^{\frac{1}{3}}$, and thus the
proton lifetime $\tau_p$ is proportional to
$\lambda^2/(\lambda^{\prime})^{\frac{2}{3}}$. This indicates that
larger $\lambda$ values and smaller $\lambda^\prime$ values help avoid
the proton decay bound.

\begin{figure}[t]
\begin{center}
\subfigure[${\cal A} (p\to K^+ \overline{\nu}_\tau)$]
 {\includegraphics[clip, width = 0.46 \textwidth]{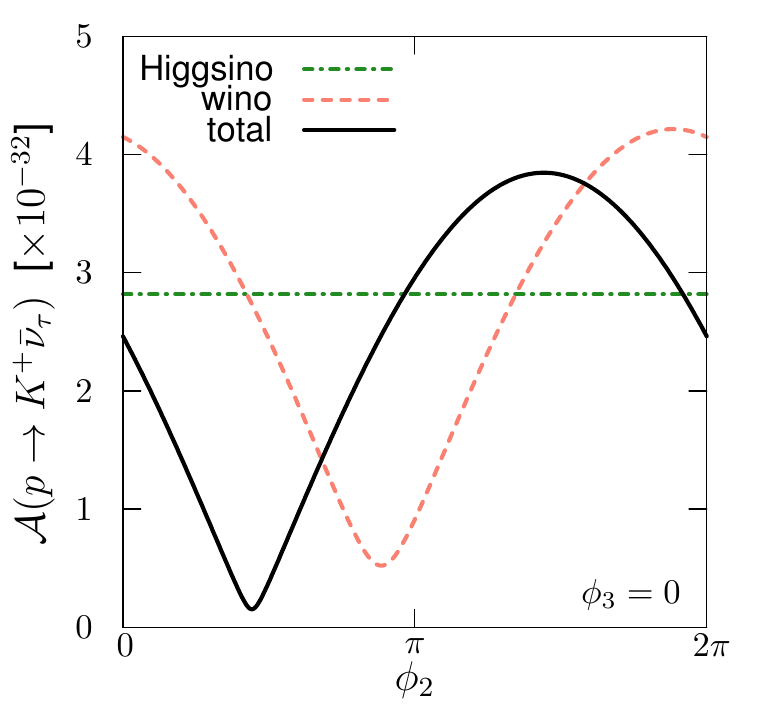}
 \label{fig:ampl}}
\subfigure[$\tau (p \to K^+ \overline{\nu}_i)$]
 {\includegraphics[clip, width = 0.51 \textwidth]{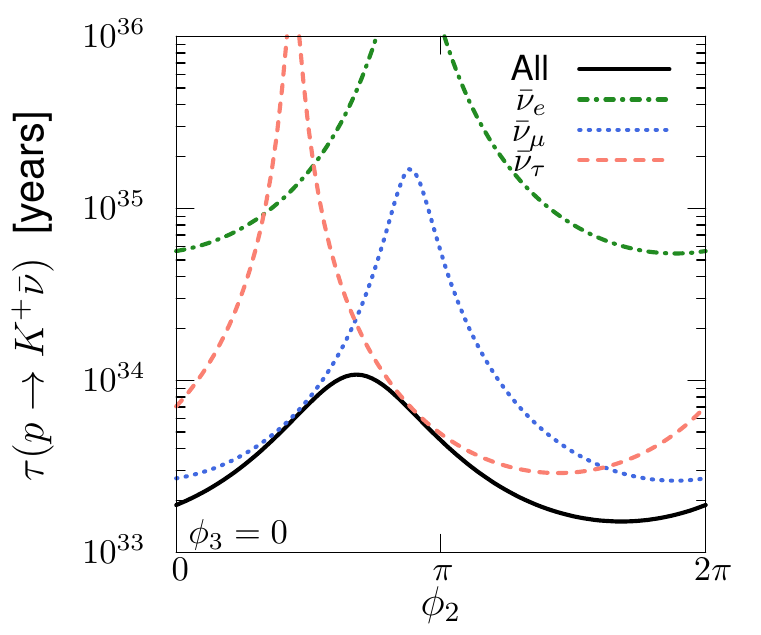}
 \label{fig:partlife}}
\caption{\it (a): The absolute value of the contributions to the decay
 amplitude of the $p\to K^+ \overline{\nu}_\tau$ channel as functions
 of $\phi_2$. The red dashed, green dash-dotted and black solid lines
 represent the absolute values of the wino, Higgsino, and total
 contributions, respectively. (b): The phase dependences of the lifetimes
 for the different $p\to K^+ \overline{\nu}$ decay modes. The green dash-dotted, blue dotted and red dashed
 lines represent the first-, second-, and third-generation neutrino
 decay modes, respectively, and the black solid line shows the total
 lifetime. In both figures, we set $\phi_3 = 0$, and take the parameter
 point indicated by the star $(\bigstar)$ in Fig.~\ref{fig:diskr1}. }
\label{fig:pknuphi2}
\end{center}
\end{figure}

To show the phase dependence of these contributions more clearly, we show in
Fig.~\ref{fig:ampl} each contribution to the decay amplitude of
the $p\to K^+ \overline{\nu}_\tau$ channel as a function of $\phi_2$
with $\phi_3$ fixed to be $\phi_3 = 0$. The red dashed, green
dash-dotted and black solid lines represent the absolute values of the
wino, Higgsino, and total contributions, respectively. We take the
parameter point indicated by the star $(\bigstar)$ in
Fig.~\ref{fig:diskr1} below. This figure shows that the wino contribution can
vary by almost an order of magnitude, while the size of the Higgsino
contribution remains constant. These contributions are comparable, and
thus a significant cancellation can occur. As a result, the total amplitude
varies by more than an order of magnitude. The wino contribution is
minimized at $\phi_2 \simeq 0.89\pi$, while the total amplitude is
minimized at $\phi_2 \simeq 0.44\pi$. This mismatch is due to the
Higgsino contribution.

In Fig.~\ref{fig:partlife} we show the phase dependence of the lifetime
of each $p\to K^+ \overline{\nu}$ decay mode with the same parameter set. The green dash-dotted,
blue dotted and red dashed lines represent the first-, second-, and
third-generation neutrino decay modes, respectively, while the black
solid line shows the total lifetime. We see that the lifetimes of the
$\overline{\nu}_e$ and $\overline{\nu}_\mu$ modes, which are induced
by wino exchange only, are maximized at $\phi_2 \simeq 0.89\pi$,
which deviates from the point where $\tau (p \to K^+
\overline{\nu}_\tau)$ is maximized. Due to this deviation, the phase dependence of
the total lifetime is much smaller than that of each partial lifetime,
but still it can change by an ${\cal O} (1)$ factor.

\begin{figure}[t]
\begin{center}
\subfigure[$\tau (p\to K^+ \overline{\nu})$]
 {\includegraphics[clip, width = 0.46 \textwidth]{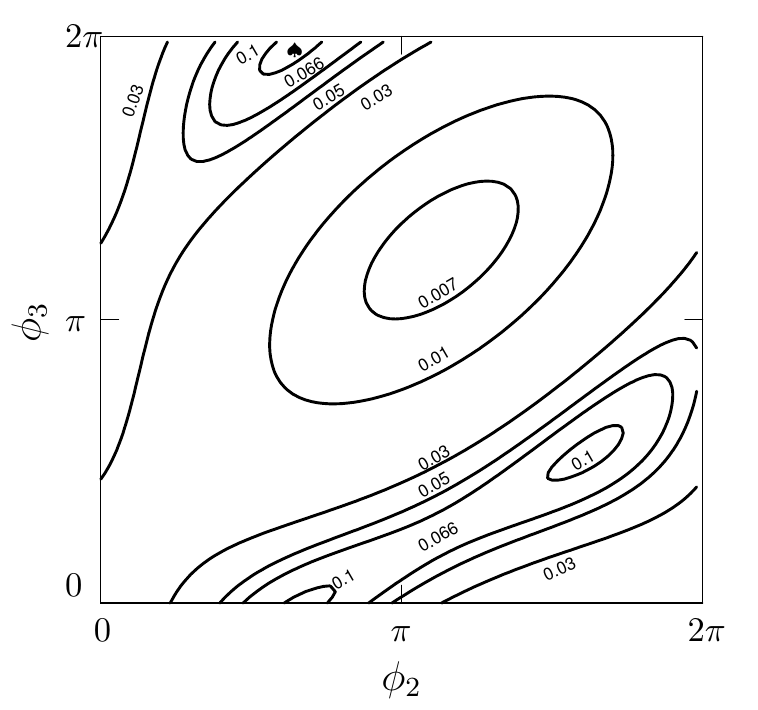}
 \label{fig:pknuphase}}
\subfigure[Nucleon decay lifetimes]
 {\includegraphics[clip, width = 0.51 \textwidth]{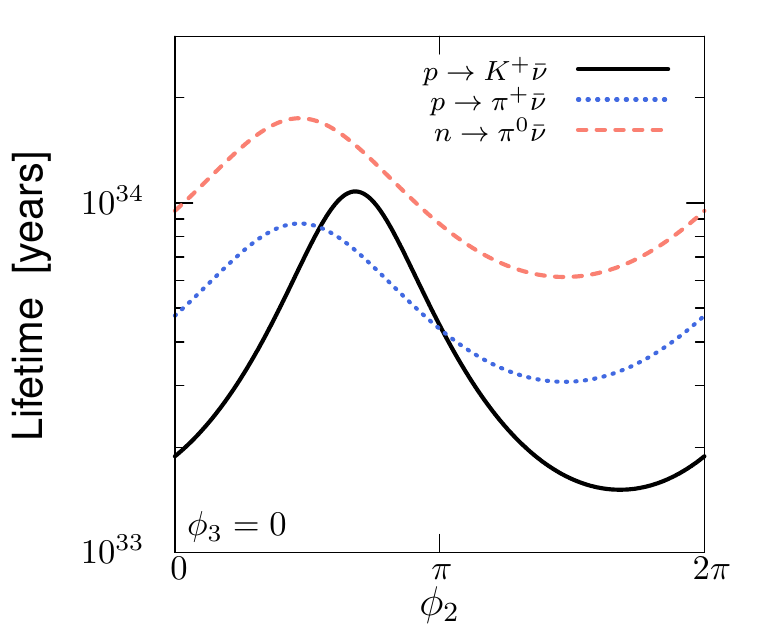}
 \label{fig:nucleon}}
\caption{\it (a): Contour plot for the proton decay lifetime in units of
 $10^{35}$ years. The area within the 0.066 contour satisfies the current experimental bound.
 The peak lifetime is denoted by $\spadesuit$. (b): Lifetimes of the nucleon decay modes as
functions of $\phi_2$. Calculated for for the reference point indicated by a star $(\bigstar)$
in Fig.~\protect\ref{fig:diskr1}.}
\label{fig:phasedep}
\end{center}
\end{figure}

In Fig.~\ref{fig:pknuphase}, we show a contour plot for the proton decay
lifetime in units of $10^{35}$ years in the $\phi_2$--$\phi_3$ plane, using
the same parameter set as in Fig.~\ref{fig:pknuphi2}. We find that
the proton lifetime exceeds the current experimental bound, $\tau (p\to
K^+ \overline{\nu}) > 6.6\times 10^{33}$~yrs \cite{Takhistov:2016eqm,
Abe:2014mwa}, in a significant area of the phase space shown by the contour labeled
0.066. The peak lifetime is marked in the upper part of the figure by a spade.

Although the $p\to K^+ \overline{\nu}$ modes may be suppressed for certain
values of the phases, other decay
modes that depend on the same phases are not suppressed in the same way.  The other
decay modes that could restrict the parameter space are $p\to
\pi^+\overline{\nu}$ and $n\to \pi^0 \overline{\nu}$. The Wilson
coefficients for these proton decay modes are quite similar to those
that generate $p\to K^+\overline{\nu}$, and depend on exactly the same
combination of SUSY parameters.  The differences in the calculations of their
lifetimes come from their different dependences on CKM matrix
elements. The $p\to \pi^+ \overline{\nu}$ and $n\to \pi^0
\overline{\nu}$ modes are suppressed relative to the $p\to K^+ \overline{\nu}$ modes
by off-diagonal components of the CKM matrix. Moreover, the experimental constraints on
these modes are weaker:
$\tau (p\to \pi^+ \overline{\nu}) > 3.9 \times 10^{32}$~yrs and $n\to \pi^0
\overline{\nu} > 1.1\times 10^{33}$~yrs \cite{Takhistov:2016eqm,
Abe:2013lua}, so these decay modes are less restrictive on the parameter
space. To ensure that these modes are not problematic, in
Fig.~\ref{fig:nucleon}, we show the lifetimes of these decay modes as
functions of $\phi_2$ for the same parameter set as in
Fig.~\ref{fig:pknuphi2}. We find that, although the $p\to \pi^+
\overline{\nu}$ mode can be dominant, it is still above the present
experimental limit. The $n\to \pi^0 \overline{\nu}$ is always
sub-dominant, and again exceeds the current bound. We also note that the
$p\to \pi^+ \overline{\nu}$ and $n\to \pi^0 \overline{\nu}$ modes
exhibit the same phase dependence, since they are related to each other
through isospin symmetry.

In the following analysis, we choose the CP-violating phases so as to maximize
the $p\to K^+ \overline{\nu}$ lifetime, thereby obtaining a conservative constraint on the
super-GUT model parameter space. Although not shown in the figures below, we have verified that each allowed point also meet the experimental constraint coming from $p\to \pi^+ \overline{\nu}$ and $n\to \pi^0 \overline{\nu}$.

\section{Results}
\label{sec:results}

To appreciate the effect of choosing $M_{in} > M_{GUT}$, we begin by
reviewing briefly some results for the CMSSM with $M_{in} = M_{GUT}$. We
note that we use here the {\tt FeynHiggs 2.11.3} code~\cite{FH} to
compute the Higgs mass. Previously we used {\tt FeynHiggs 2.10.0}, and
we note that due to a bug fix, the new version yields a significant
change in $m_h$ at large positive $A_0$~\footnote{Note that our sign
convention for $A_0$ is opposite that found in many public codes such as
{\tt SoftSusy}~\cite{softsusy}.}. A large value of $A_0/m_0$ is
necessary to obtain the correct relic density along the stop-coannihilation strip \cite{stop,eoz}, where the lighter stop and
neutralino LSP are nearly degenerate in mass. For $A_0/m_0 \gtrsim 2$,
we find that {\tt FeynHiggs 2.11.3} results in a $\simeq 1.5$~GeV drop in the
value of $m_h$ relative to the previous result, necessitating a
lower value of $A_0/m_0$. However, for $A_0/m_0 \lesssim 2$, the stop
strip is no longer present. On the other hand, the effect of updating
{\tt FeynHiggs} on $m_h$ at large negative $A_0/m_0$ is less
pronounced. We further note that our calculation of the proton lifetime
here is also updated with bug-fixes.

\subsection{CMSSM update}
\label{sec:cmssm}

In view of the proton lifetime constraint, which favours larger
sparticle masses, we consider here the possibilities that the correct
relic density of neutralino dark matter is obtained either in the
focus-point strip \cite{fp2,fp} or the stop-coannihilation strip
\cite{stop}, updating the results found in \cite{Ellis:2015rya}.
We use SSARD \cite{SSARD} to compute the particle mass spectrum, the
dark matter relic density, and proton lifetimes. The discussion of the
proton lifetime in Section~\ref{pdecay} motivates us to focus on
relatively small values of $\tan \beta$.  For larger values of $\tan
\beta$, the proton lifetime becomes smaller than the current experimental bound,
and minimal supersymmetric SU(5) is not viable. For the CMSSM cases with
$M_{in} = M_{GUT}$, we have set $c=0$ and taken $M_{H_C}$ from
Eq.~\eqref{eq:matchmhc}. 

\begin{figure}[htb!]
\begin{minipage}{8in}
\includegraphics[height=3.in]{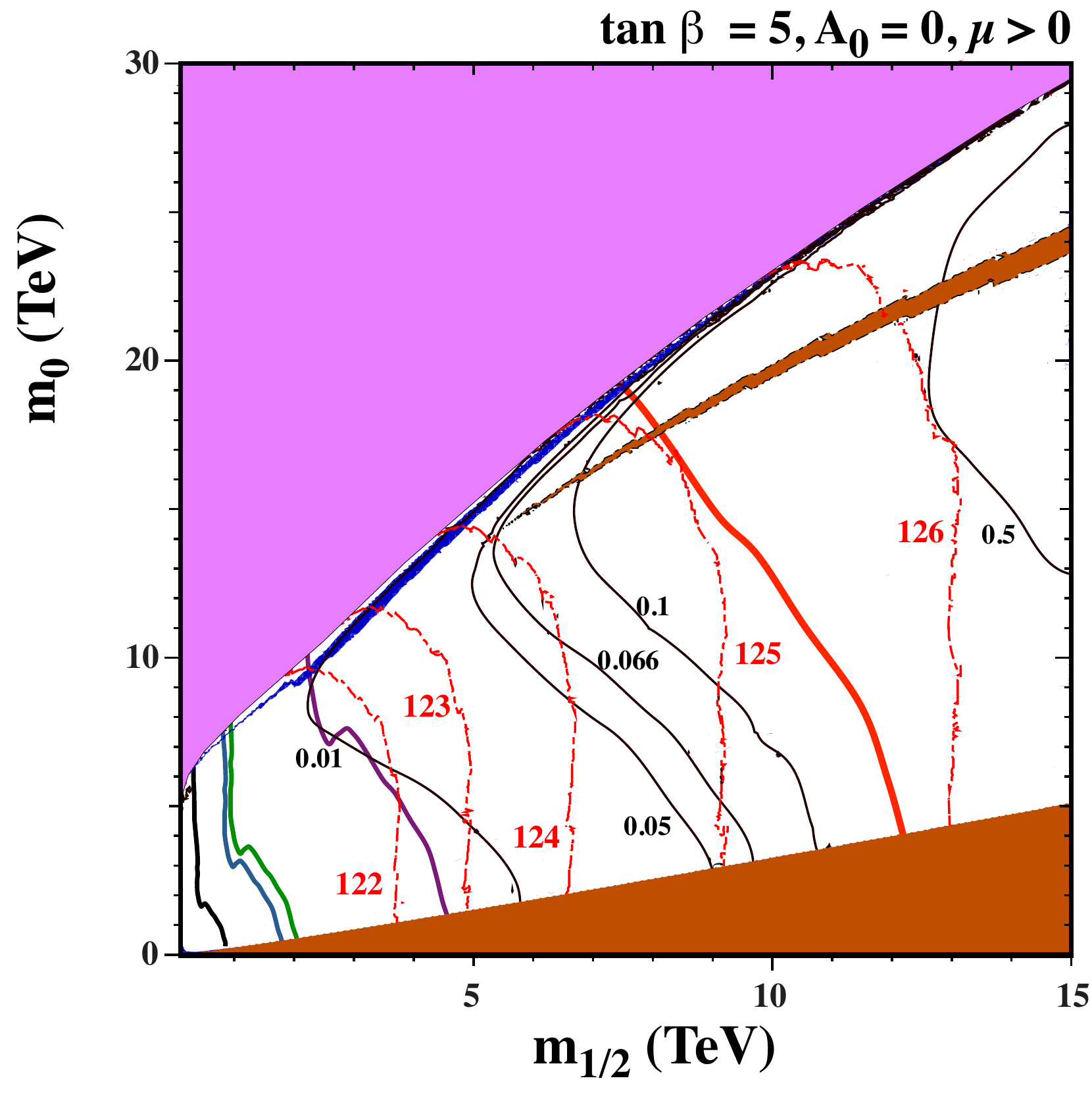}
\hspace*{-0.17in}
\includegraphics[height=3.in]{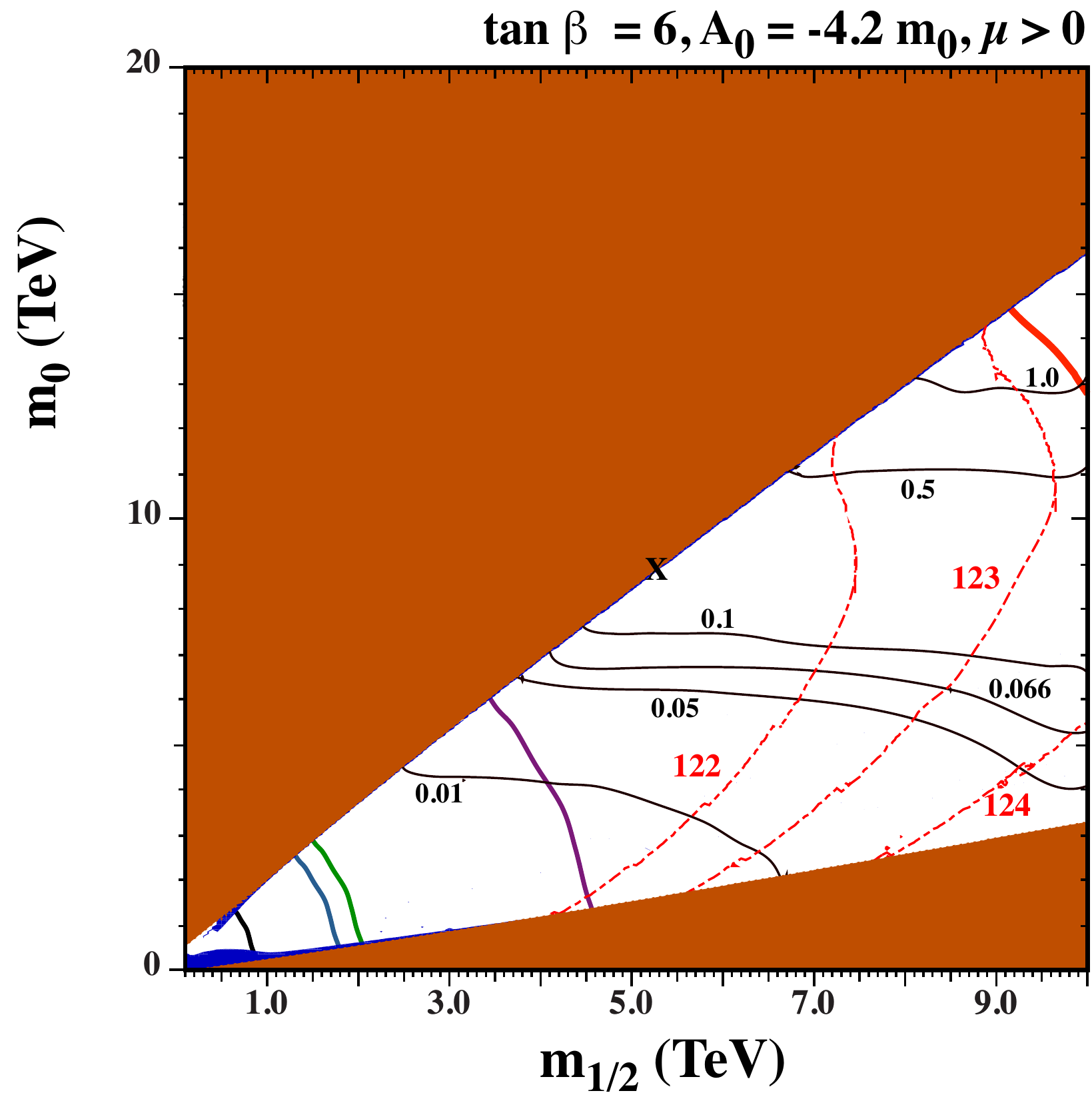}
\hfill
\end{minipage}
\begin{minipage}{8in}
\includegraphics[height=3.in]{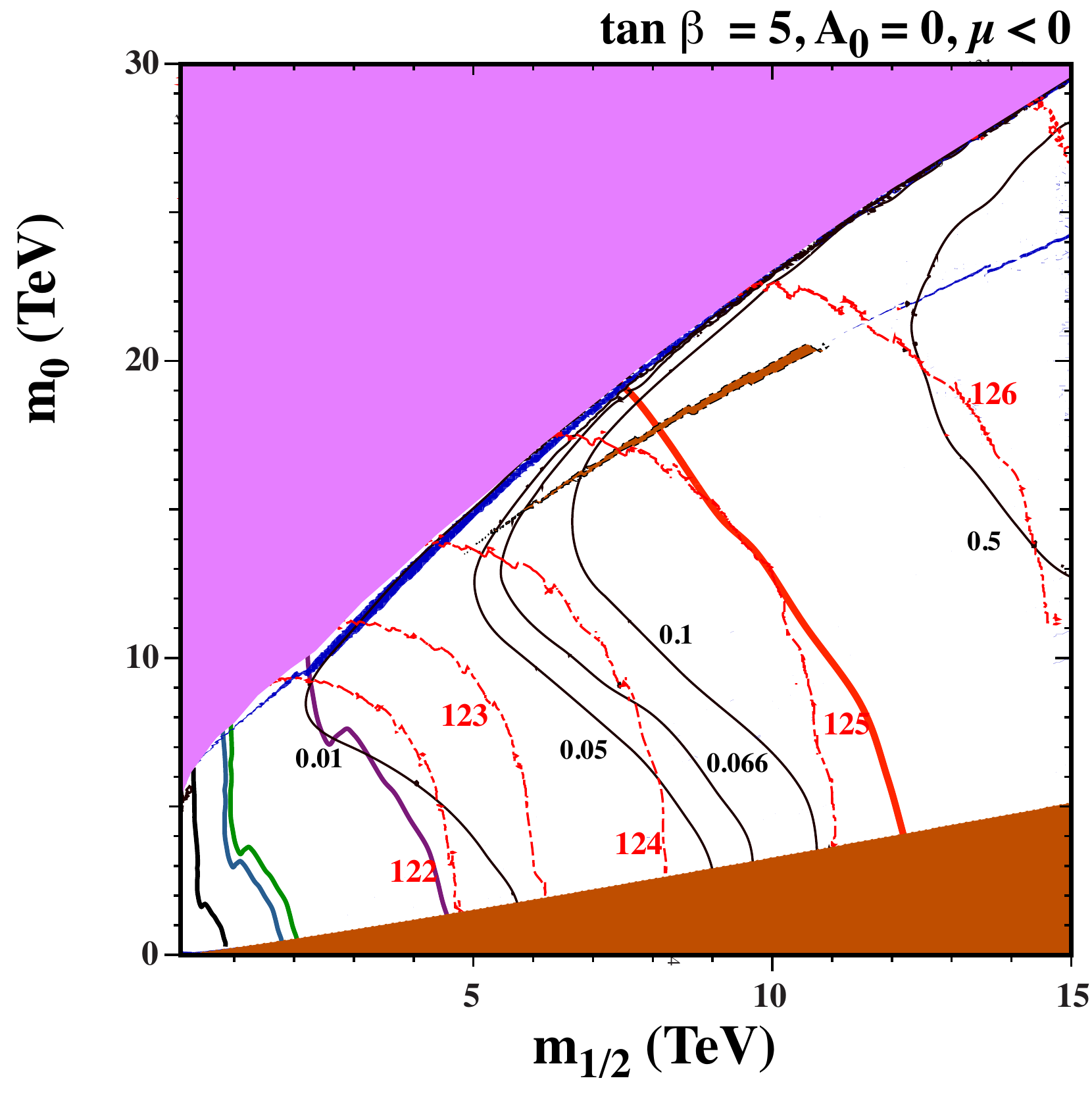}
\hspace*{-0.17in}
\includegraphics[height=3.in]{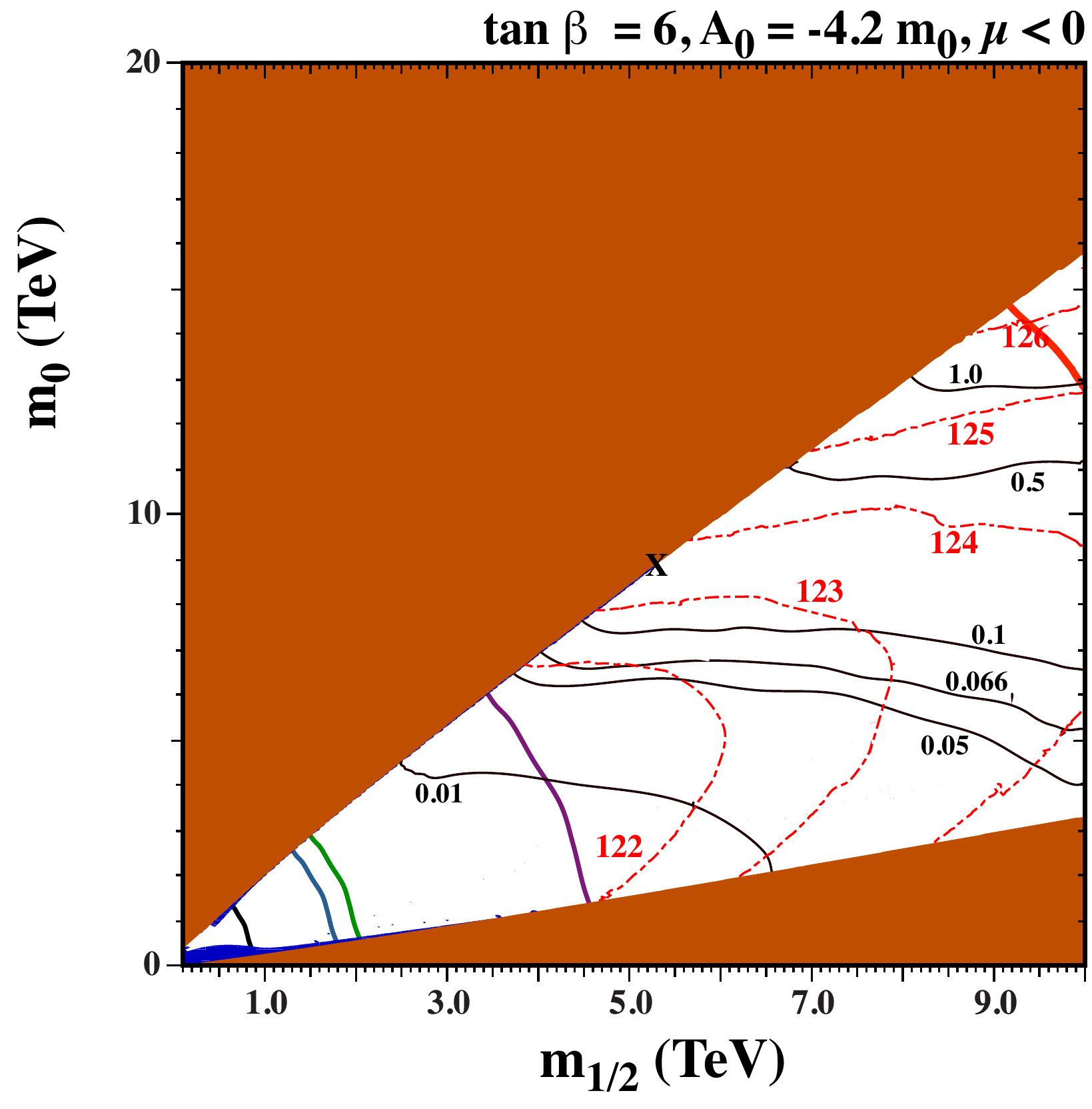}
\end{minipage}
\caption{
{\it
Sample CMSSM $(m_{1/2}, m_0)$ planes showing the focus-point strip for
 $\tan \beta = 5$ and $A_0 = 0$ (left) with $\mu > 0$ (upper) and $\mu <
 0$ (lower), and the stop coannihilation strip with $\tan \beta = 6$ and
 $A_0 = -4.2 \, m_0$ (right). In the light mauve shaded regions, it is
 not possible to satisfy the electroweak symmetry breaking 
 conditions. In the brown shaded regions, the LSP is charged and/or
 colored. The dark blue shaded strips show the areas where $0.06 <
 \Omega_\chi h^2 < 0.2$ in the left panels and the further enlarged
 range of $0.01 < \Omega_\chi h^2 < 2.0$ in the right panels. The red
 dot-dashed contours indicate the Higgs mass, labeled in GeV, and the
 solid black contours indicate the proton lifetime in units of
 $10^{35}$~yrs. The bold solid black, blue, green, purple, and red lines
 in each panel are current and future limits from the LHC at 8 TeV, $300$
 and $3000$~fb$^{-1}$ at 14 TeV, 3000~fb$^{-1}$ with the HE-LHC at
 33~TeV, and 3000~fb$^{-1}$ with the FCC-hh at 100~TeV, respectively,
 taken from the analysis of \protect{\cite{Buchmueller:2015uqa}}.
}}
\label{fig:CMSSM}
\end{figure}

In Fig.~\ref{fig:CMSSM}, we show four CMSSM $(m_{1/2},m_0)$ planes
displaying the focus-point (left) and stop-coannihilation (right) relic
density strips for the two choices of the sign of $\mu$. Higgs mass
contours are shown as red dot-dashed curves labeled by $m_h$ in GeV in
1~GeV intervals starting at 122~GeV. In the left panels, we choose $A_0
= 0$ \footnote{As we discussed in Section~\ref{sec:matchingcond}, if we
assume the minimal SU(5) GUT with the universality condition
\eqref{eq:inputcond}, then the $B$-term matching condition restricts
$A_0$ via Eq.~\eqref{alimit}. This constraint can, however, be evaded if
we relax the universality condition \eqref{eq:inputcond} (for $m_\Sigma$
in particular) or consider non-minimal Higgs content. With these
possibilities in mind, we do not take the condition \eqref{alimit} into
account in Section~\ref{sec:cmssm}, which allows the choice $A_0 = 0$. 
} with $\mu > 0$ (top) and $\mu < 0$ (bottom). For this choice of $A_0$,
there is a relatively minor effect on $m_h$ due to the updated version
of {\tt FeynHiggs}. The light mauve shaded region in the parts of the
left panels with large $m_0/m_{1/2}$ are excluded because there are no
solutions to the EWSB conditions: along this boundary $\mu^2 = 0$. Just
below the regions where EWSB fails, there are narrow dark blue strips
where the relic density falls within the range determined by CMB and
other experiments~\cite{Planck15}~\footnote{Since the relic density of
dark matter is now determined quite accurately ($\Omega_\chi h^2  =
0.1193 \pm 0.0014$), for the purpose of visibility we display expanded
strips for which the relic density lies in the range $[0.06,
0.20]$.}. These strips are in the focus-point region~\cite{fp,fp2}. We
note also that the brown shaded regions in the portions of the panels
with low $m_0/m_{1/2}$ are excluded because there the LSP is the lighter
charged stau lepton. The planes also feature stau-coannihilation strips
\cite{stau} close to the boundaries of these brown shaded regions. They
extend to $m_{1/2} \simeq 1$~TeV, but are very difficult to see on the
scale of this plot, even with our enhancement of the relic density
range. There are also `thunderbolt'-shaped brown shaded bands at
intermediate $m_0/m_{1/2}$ where the chargino is the LSP. There are no
accompanying chargino-coannihilation strips, as at these multi-TeV
mass scales any such strip would lie within the shaded region and is
therefore excluded. 

Contours of the proton lifetime calculated using down-type Yukawa couplings (see
the discussion in Section~\ref{sec:matchingcond}) are shown as solid
black curves that are labeled in units of $10^{35}$~yrs. The current
limit $\tau_p > 6.6 \times 10^{33}$~yrs \cite{Takhistov:2016eqm,
Abe:2014mwa} would exclude the entire area below the curve labeled
0.066. For the nominal value of $m_h = 125$~GeV, neglecting the
theoretical uncertainties in the calculation of $m_h$, we see that in
the upper left plane of Fig.~\ref{fig:CMSSM} the Higgs contour
intersects the focus-point region where $\tau_p \approx 5 \times
10^{33}$~yrs, very close to the experimental limit. Much of the
focus-point strip in this figure may be probed by future proton decay
experiments. Changing the sign of $\mu$ has almost no effect on the
proton lifetime, as seen in the lower left panel of
Fig.~\ref{fig:CMSSM}, but the calculated Higgs mass is smaller by $\sim
1$~GeV, which is less than the uncertainty in the {\tt FeynHiggs}
calculation of $m_h$. 

In the right panels of Fig.~\ref{fig:CMSSM}, we have chosen large
negative $A_0/m_0 = -4.2$ and $\tan \beta = 6$, which allows a
sufficiently heavy Higgs and a viable stop strip. There are now brown
shaded regions in the upper left wedges of the planes where the stop is
the LSP (or tachyonic). Though it is barely visible, there is a stop
strip that tracks that boundary \footnote{In this case, and in the
super-GUT cases to follow, we have further extended the range on
$\Omega_\chi h^2$ to [0.01,2.0]. Otherwise the thickness of the strips
which are typically 10--50~GeV would be pixel thin for the range of
masses shown.}. Since we have taken an enhanced range for the relic
density the blue strip continues to the edge of the plot. In reality,
however, the stop strip ends~\cite{eoz} at the position marked by the
{\bf X} in the figure. We see that, for $\mu > 0$, the stop strip ends
when $m_h < 122$~GeV, whereas for $\mu < 0$ the strip ends when $m_h
\approx 123.5$~GeV, both of which are acceptable given the uncertainty
in the calculation of $m_h$. At the endpoint, which occurs at $(m_{1/2},
m_0) \simeq (5.2,8.8)$~TeV, the proton lifetime is approximately
$2\times 10^{34}$~yrs. Had we chosen a smaller value of $|A_0/m_0|$, the
stop strip would have extended to higher $m_h$. For example, for $\mu <
0$, the stop strip extends to 125~GeV for $A_0/m_0 = -3.5$ and the
endpoint is found at (5.1,11.3) TeV.

In all of the cases shown in Fig.~\ref{fig:CMSSM}, the favored parameter
regions predict the masses of supersymmetric particles to be in the
multi-TeV range. For example, as the gluino mass is $\simeq
2\times m_{1/2}$, it is expected to be as large as $\simeq 10$~TeV,
which is well above the LHC reach \cite{ATLAS20,
CMS20}. To see the current and future limits on the CMSSM parameter
space from the LHC and future hadron colliders such as the 33~TeV HE-LHC
option and the Future Circular Collider (FCC) \cite{Golling:2016gvc} which
aims at 100~TeV proton-proton collisions, we show the limits from LHC
at 8 TeV, and sensitivities with $300$ and $3000$~fb$^{-1}$ with the LHC at 14 TeV, 3000~fb$^{-1}$
with the HE-LHC at 33~TeV, and 3000~fb$^{-1}$ with the FCC-hh at 100~TeV as the
bold solid black, blue, green, purple, and red lines in each panel in
Fig.~\ref{fig:CMSSM}, respectively, following the analysis given in
Ref.~\cite{Buchmueller:2015uqa}. As we see, the parameter region in
which the proton decay bound is evaded is far beyond the reach of the
LHC, but may be probed at the 100 TeV collider. We further note that,
while the stop-coannihilation region shown may not be fully probed at 33 TeV,
the 100 TeV reach clearly extends beyond the stop endpoint marked by the {\bf X}.
On the other hand, the focus-point region is seen to extend beyond the 100 TeV reach.

\subsection{Super-GUT CMSSM}

\begin{figure}[htb!]
\begin{minipage}{8in}
\includegraphics[height=3.3in]{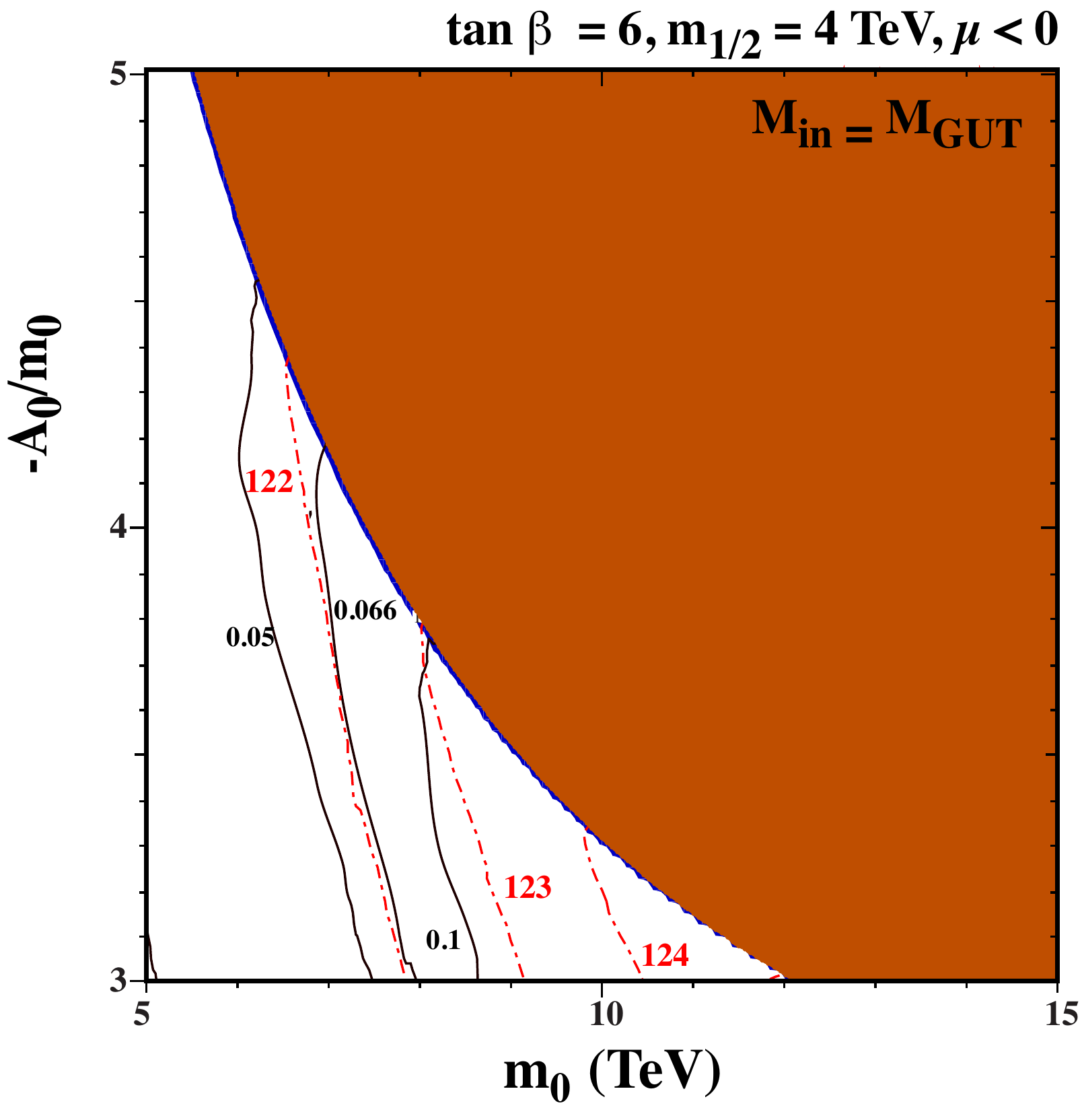}
\includegraphics[height=3.3in]{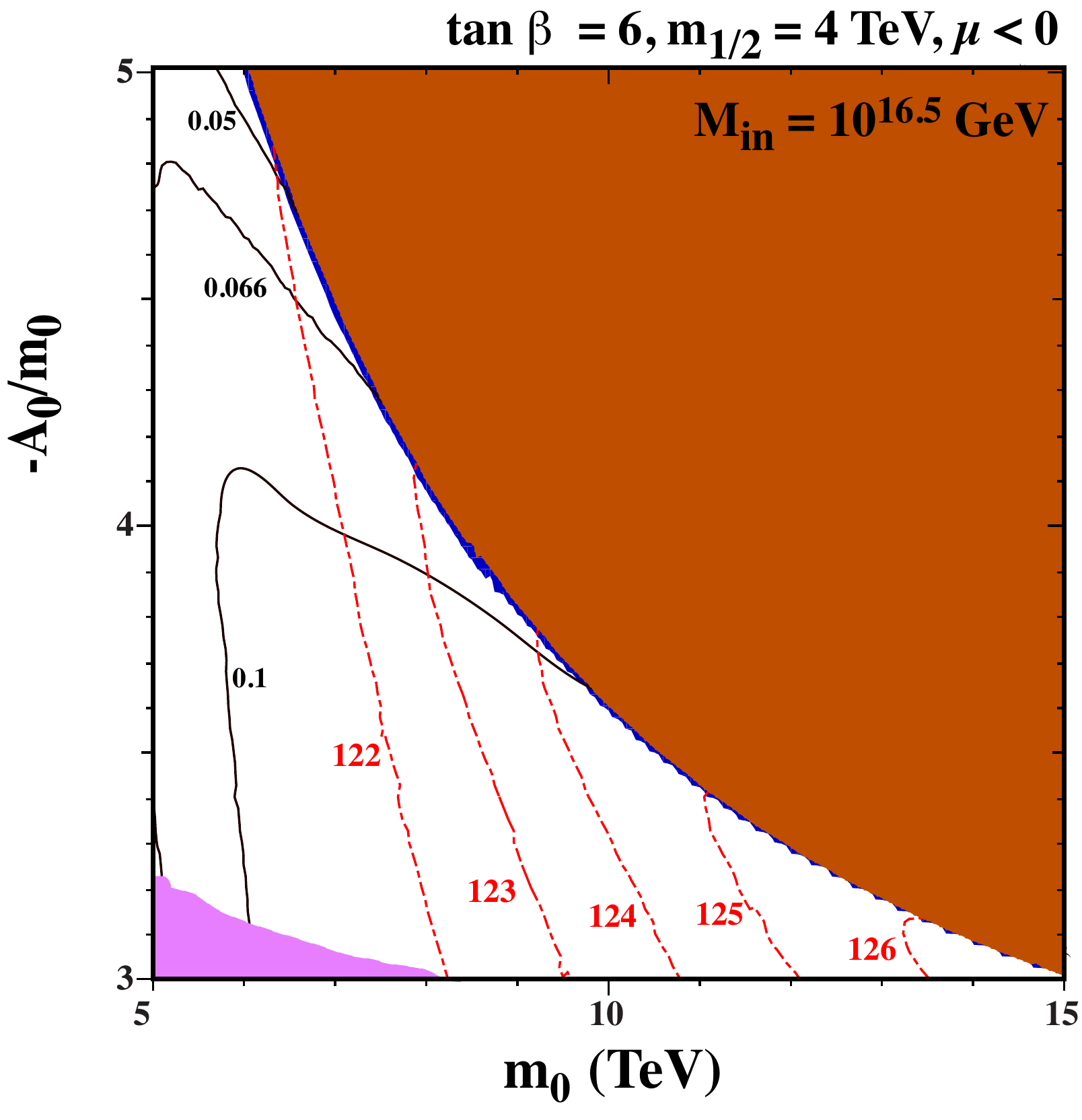}
\hfill
\end{minipage}
\begin{minipage}{8in}
\includegraphics[height=3.3in]{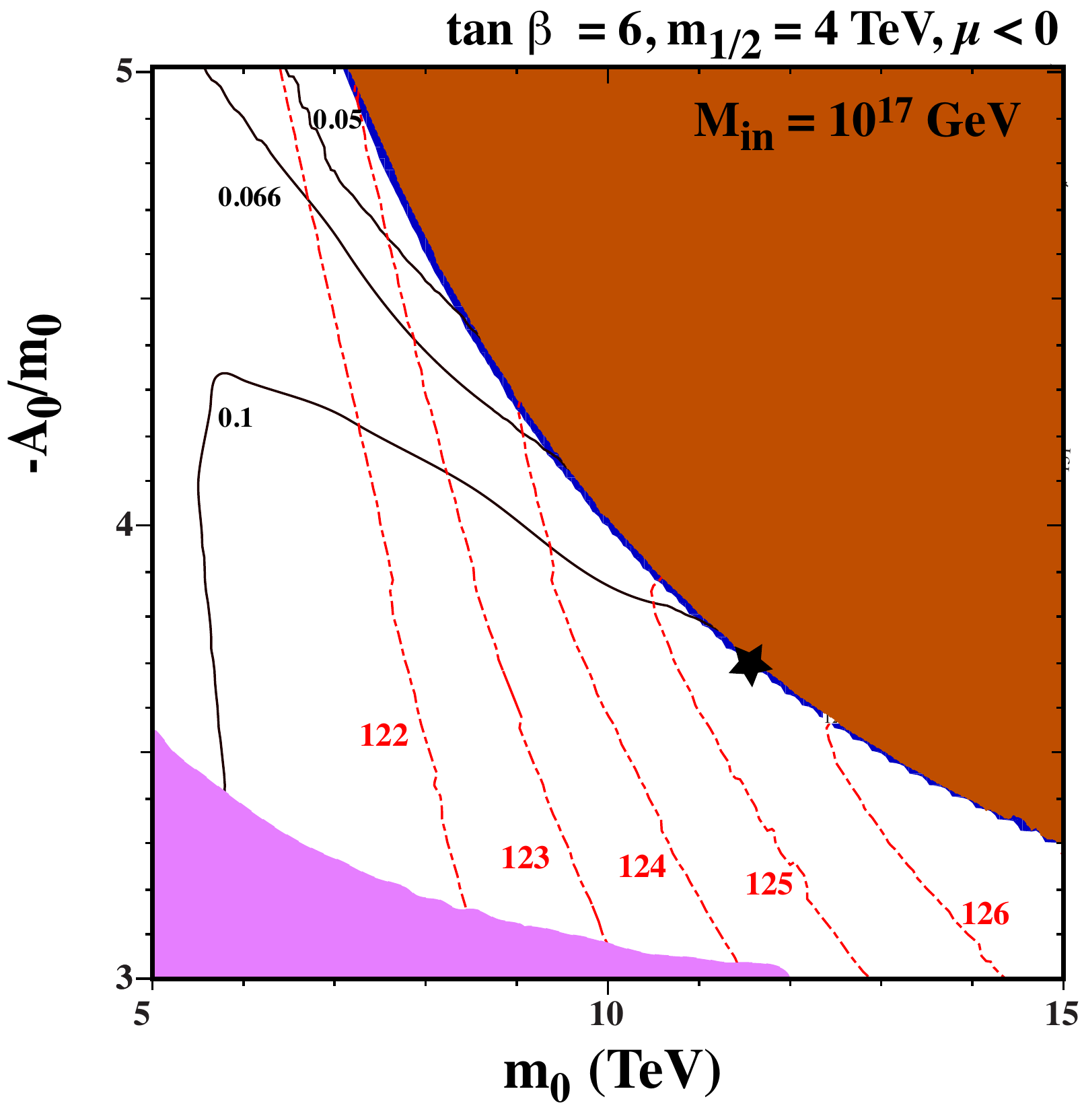}
\includegraphics[height=3.3in]{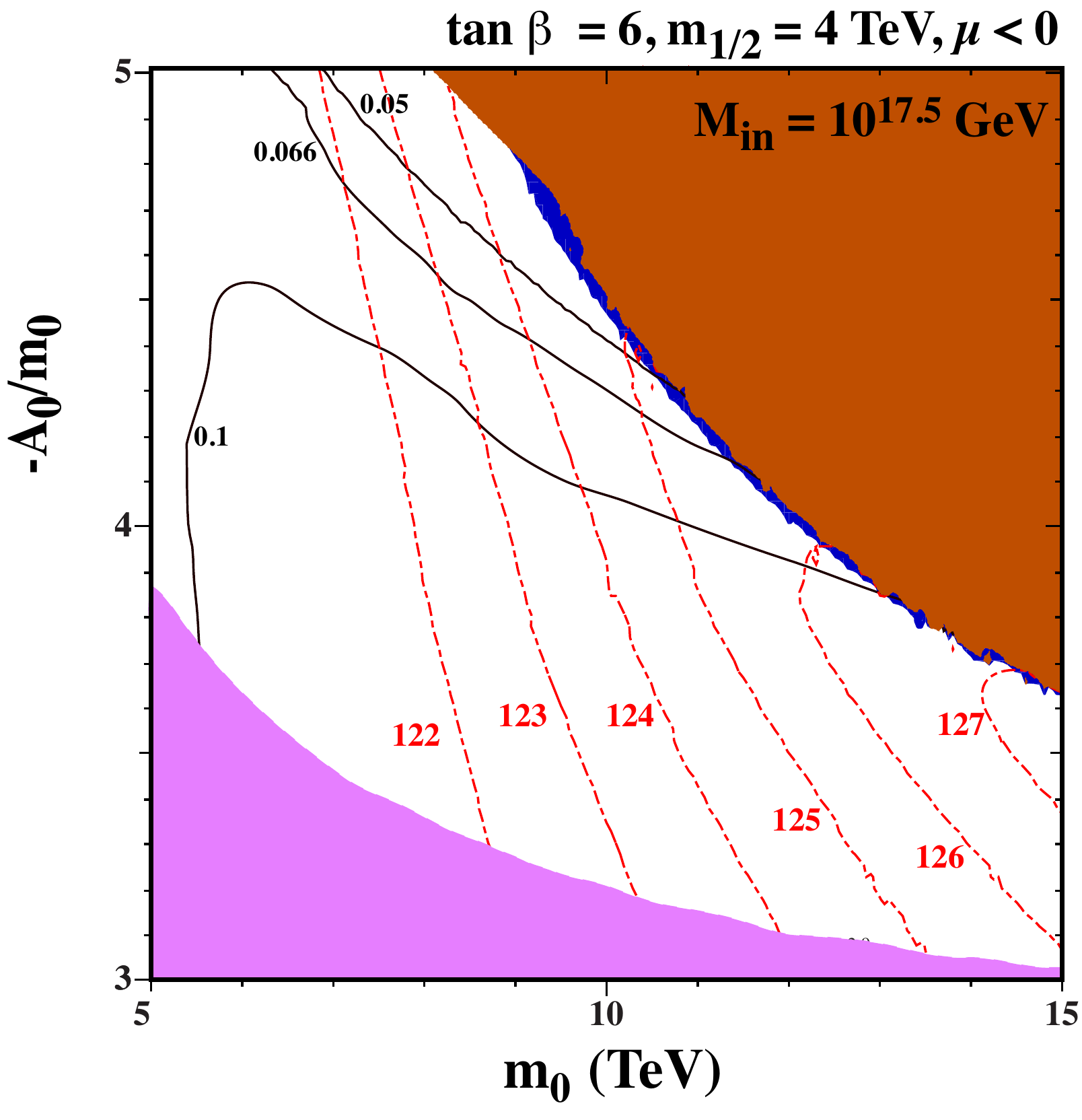}
\end{minipage}
\caption{
{\it
Super-GUT CMSSM $(m_{0}, A_0/m_0)$ planes for $\tan \beta = 6$ and $\mu
 < 0$. The values of $M_{in}$ are $M_{GUT}$, $10^{16.5}$, $10^{17}$ and
 $10^{17.5}$~GeV, as indicated. In each panel, we have fixed $m_{1/2} =
 4$~TeV, $\lambda = 0.6$ and $\lambda^\prime = 0.0001$. In the light
 mauve shaded regions, it is not possible to satisfy the matching
 condition for $B$. In the brown shaded regions, the LSP is the
 stop. The dark blue shaded regions show the areas where $0.01 <
 \Omega_\chi h^2 < 2.0$.  The red dot-dashed contours indicate the Higgs
 mass, labeled in GeV, and the solid black contours indicate the proton
 lifetime in units of $10^{35}$~yrs. 
}}
\label{fig:diskr1}
\end{figure}

As we discussed earlier, the super-GUT scenario introduces several new
parameters, making a complete analysis quite cumbersome. In addition to
the CMSSM parameters, we must specify the input universality scale
$M_{in}$ and the values of the two GUT couplings $\lambda$ and
$\lambda'$. In order to understand better the parameter space of the
super-GUT models, we begin by considering $(m_0,A_0/m_0)$ planes for
fixed $m_{1/2}, \tan \beta, \lambda$, and $\lambda'$ and several choices
of $M_{in}$, as shown in Fig.~\ref{fig:diskr1}.

In Fig.~\ref{fig:diskr1}, we have fixed $\tan \beta = 6, m_{1/2} =
4$~TeV, $\lambda = 0.6$ and $\lambda^\prime = 10^{-4}$ with $\mu <
0$. We have chosen $M_{in} = M_{GUT}$, $10^{16.5}$, $10^{17}$ and
$10^{17.5}$~GeV in the upper left, upper right, lower left and lower
right panels, respectively. In each panel, the contours for $m_h$ and
$\tau_p$ are drawn using the same line styles as in the previous figure.
The brown shaded regions at large $m_0$ and $- A_0/m_0$ are excluded
because they contain a stop LSP (or tachyonic stop), and the stop relic
density strip tracks this boundary. Because $m_{1/2}$ is fixed, there is
no endpoint of the strip within the parameter ranges shown, and the
lightest neutralino is an acceptable LSP everywhere along the blue strip
(remembering that the thickness of the strip is exaggerated for clarity).
For $M_{in} > M_{GUT}$, there is a mauve shaded region at small $m_0$
and $- A_0/m_0$ that grows in size as $M_{in}$ is increased. In this
region, the $B$ matching condition \eqref{eq:matchingb} is violated, and
there is no solution to \eqref{alimit}~\footnote{For $M_{in} = M_{GUT}$,
the region excluded is $|A_0| \lesssim 2.8 m_0$, which is below the
range displayed in the Figure.}.

When $M_{in} = M_{GUT}$ with the  parameters adopted in
Fig.~\ref{fig:diskr1}, the Higgs mass prefers smaller values of
$|A_0/m_0|$ and larger values of $m_0$. In the portion of the strip
where $m_h > 123$~GeV according to {\tt FeynHiggs} (which is consistent
with the experimental measurement), the proton lifetime is $> 10^{34}$
yrs. As $M_{in}$ is increased, we see that the stop LSP region moves to
larger $m_0$ and $|A_0/m_0|$, while low values of $|A_0/m_0|$ are
excluded because of the $B$ matching condition. For $m_h = 125$ GeV, the
allowed values of $m_0$ and $|A_0/m_0|$ increase as $M_{in}$ is
increased. For very large $M_{in}$, we see that the intersection of the
$m_h$ contour with the stop strip occurs at lower $\tau_p$ and for
$M_{in} = 10^{17.5}$~GeV, the intersection point occurs below the
current experimental bound. The star $(\bigstar)$ in the
lower left panel with $M_{in} = 10^{17}$ GeV, is a benchmark we used
in Section~\ref{pdecay} to discuss the choice of phases. At this point,
which is located at $m_0 = 11.6$ TeV and $A_0/m_0 = -3.7$, we must take
$c = -0.0095$ in Eq.~\eqref{eq:SigmaWW} in order to obtain $\lambda =
0.6$ with $\lambda^\prime = 10^{-4}$ and we find that the Higgs mass is
$m_h = 125.6$ GeV and $\tau_p \approx 10^{34}$ yrs. As shown in
Fig.~\ref{fig:pknuphase}, this lifetime requires phases $(\phi_2,
\phi_3) = (0.64, 1.96)\pi$.  If the phases vanish, the lifetime drops by
a factor of about 5 to $\tau_p = 1.9 \times 10^{33}$ yrs. The mass
spectrum at this point is shown in Table~\ref{table}. As can be seen,
the gluino mass is $\simeq 2m_{1/2}\simeq 8$~TeV, which is within the
reach of the 100~TeV collider \cite{Golling:2016gvc}. On the other hand,
squark masses are $\gtrsim 10$~TeV, and thus it may be difficult to
discover squarks even at the 100~TeV collider. 

\begin{table}[htb!]
\begin{center}
\begin{tabular}{cc|cc} 
\hline
\hline
Particle   &  Mass [TeV]     & Particle & Mass [TeV] \\ 
\hline
\hline         
${\chi_1^0}$        &  1.75     & $\chi_2^0$ & 3.45 \\
${\chi_3^0}$        &  12.8   & $\chi_4^0$ &12.8 \\
${\chi^\pm_1}$        &  3.45    & $\chi_2^\pm$ & 12.8\\
$h$        &  0.1256     & $H$ & 14.9 \\
$A$        &  14.9      & $\tilde{g}$ & 7.97   \\
$\tilde{e}_L$        &  11.8     & $\tilde{e}_R$ & 12.0 \\
$\tilde{\nu}_e$        &  11.8    &  &  \\
$\tilde{\tau}_1$        &  8.29     & $\tilde{\tau}_2$ & 11.8 \\
$\tilde{\nu}_\tau$        &  11.8    &  & \\
$\tilde{u}_L$        &  13.2     & $\tilde{u}_R$ & 12.9 \\
$\tilde{d}_L$        &  13.2    &  $\tilde{d}_R$  & 13.0  \\
$\tilde{t}_1$        &  1.76     & $\tilde{\tau}_2$ & 7.48 \\
$\tilde{b}_1$        &  7.34   &  $\tilde{b}_2$  & 12.9 \\
\hline
\hline
\end{tabular}
\caption{\it Particle Spectrum at the benchmark point indicated by a
 star $(\bigstar)$ in Fig.~\ref{fig:diskr1}.} 
\label{table}
\end{center}
\end{table}

The dependence of these results on $m_{1/2}$ can be gleaned from
Fig.~\ref{fig:CMSSM}. For smaller $m_{1/2}$, the Higgs mass and proton
lifetime both decrease. At higher $m_{1/2}$, we approach the endpoint of
the stop strip. For example, when $m_{1/2} = 6$~TeV, there would be no
blue strip alongside the red region (which would look similar to the
case displayed), as the relic density would exceed the Planck value even
for degenerate stops and neutralinos. The results scale as one might
expect with $\tan \beta$. At higher $\tan \beta$, the Higgs mass
increases while the proton lifetime decreases. For example, at $\tan
\beta = 7$, for the same value of $A_0/m_0$, the position of the star
when $M_{in} = 10^{17}$~GeV moves slightly to $m_0 = 11.5$ TeV, and the
Higgs mass increases to 126.1~GeV according to {\tt FeynHiggs}, but
$\tau_p$ decreases to $6.2 \times 10^{33}$ yrs. 

From the discussion in section~\ref{pdecay}, we expect that there is a
strong dependence of $\tau_p$ on $\lambda^\prime$, while little else is
affected. For example, increasing (decreasing) $\lambda^\prime$ by an
order of magnitude moves the stop-coannihilation strip of the lower left
panel of Fig.~\ref{fig:diskr1} so that the star would be at 12.1~TeV
(11.2~TeV) for $A_0/m_0$ unchanged. The Higgs mass, $m_h$, for this
shifted point is almost unchanged, 125.8~GeV (125.5~GeV), while $\tau_p$
drops by a factor of 5 (increases by a factor of 4). The dependence on
$\lambda$ is discussed in more detail below. We also checked on the
effect of changing the sign of $\mu$ and the ratio of $m_\Sigma/m_0$ for
the case considered in the lower left panel of
Fig.~\ref{fig:diskr1}. For both changes, the stop strip and proton
lifetime are barely altered. For $\mu > 0$, the Higgs mass drops
significantly. At the position of the star, the Higgs mass is 117~GeV
for $\mu > 0$. For this reason we have largely focused on $\mu < 0$ in
this paper. For $m_\Sigma/m_0 = 0.1$ the only noticeable change in the
figure is the absence of the $B$ matching constraints which is greatly
relaxed when $m_\Sigma < m_0$. We note that, for $m_\Sigma^2 = 0$ or even
negative, we are able to recover solutions with $A_0 = 0$. However, when
$M_{in} > M_{GUT}$, one does not find a a focus-point region as
discussed previously \cite{emo}. 

\begin{figure}[htb!]
\begin{minipage}{8in}
\includegraphics[height=3.3in]{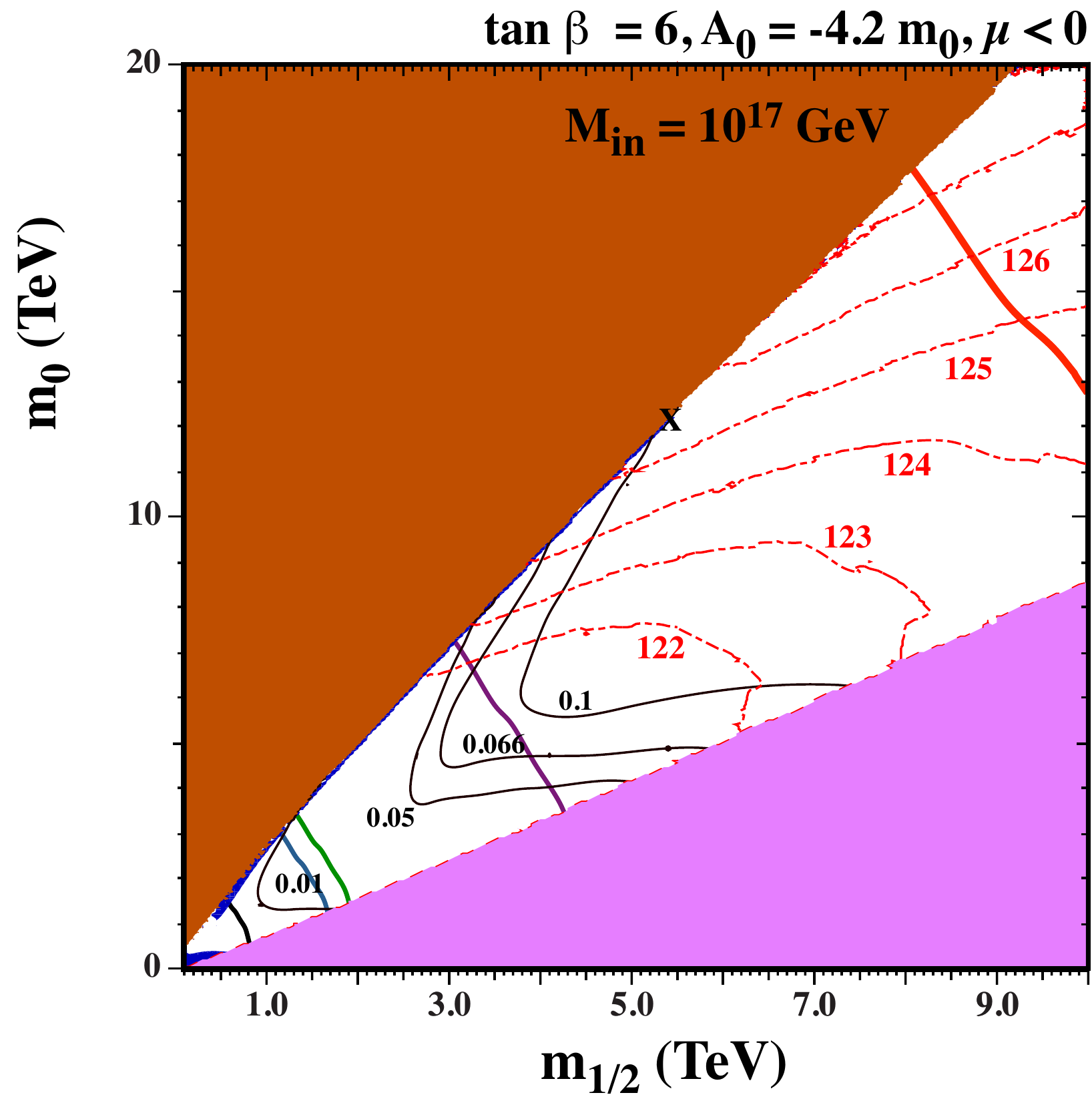}
\includegraphics[height=3.3in]{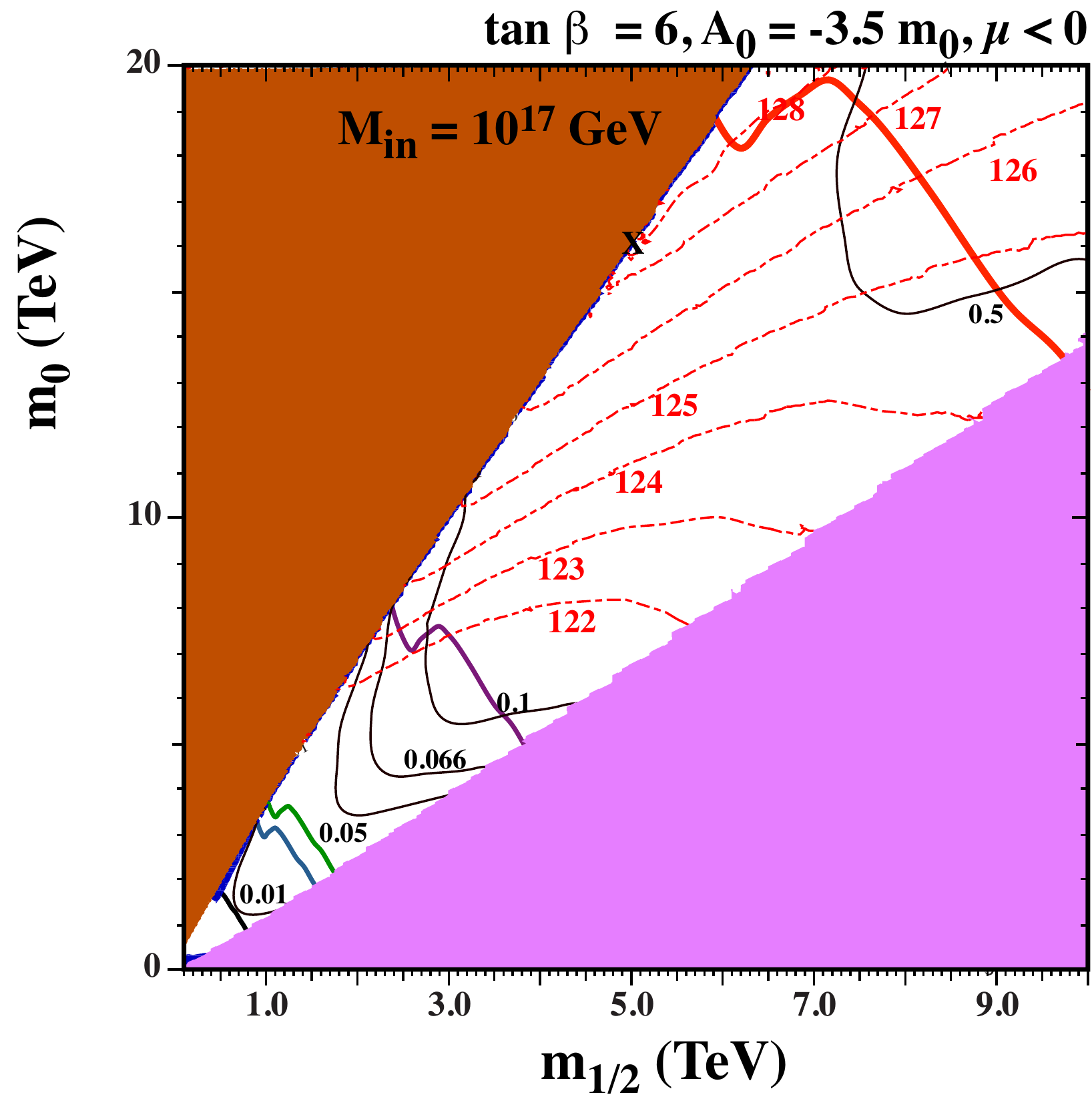}
\hfill
\end{minipage}
\caption{
{\it
Super-GUT CMSSM $(m_{12}, m_0)$ planes for $M_{in} = 10^{17}$~GeV, $\tan \beta = 6$ and
$\mu < 0$, for $A_0/m_0 = -4.2$ (left) and $-3.5$ (right). In each panel,
we have fixed $\lambda = 0.6$ and $\lambda^\prime = 0.0001$.
Shadings and contours are as in Fig. \ref{fig:CMSSM}.
The mauve shaded regions are excluded because it is not possible to satisfy the matching condition for $B$.
The $X$ marks the endpoint of the stop coannihilation strip.}}
\label{fig:superG}
\end{figure}

We next show two examples of $(m_{1/2},m_0)$ planes for $M_{in} =
10^{17}$~GeV, $\tan \beta = 6$ and $\mu < 0$, which can be compared with
the lower right panel of Fig.~\ref{fig:CMSSM}. In the left panel of
Fig.~\ref{fig:superG} we choose $A_0/m_0 = -4.2$ as in
Fig.~\ref{fig:CMSSM}. For this value of $M_{in}$, we see the appearance
of a mauve shaded region that is excluded because the $B$ matching
condition \eqref{alimit} cannot be satisfied. The $X$ located at (5.3,
12.0)~TeV again denotes the endpoint of the stop strip. This occurs when
$m_h = 125.5$~GeV and $\tau_p = 1.1 \times 10^{34}$~yrs. Thus only a
short segment of the stop strip is viable in this case. In the right
panel with $A_0/m_0 = -3.5$, we see that a larger fraction of the plane
is excluded by the failure to satisfy the $B$ matching condition. The
stop endpoint has moved to higher mass scales $(m_{1/2}, m_0) =
(5,16)$~TeV, where $m_h = 128.1$~GeV and $\tau_p = 2 \times
10^{34}$~yrs, and a larger portion of the strip is viable. In both
cases, the viable parameter points can be probed at future collider
experiments. 

\begin{figure}[htb!]
\begin{minipage}{8in}
\includegraphics[height=3.3in]{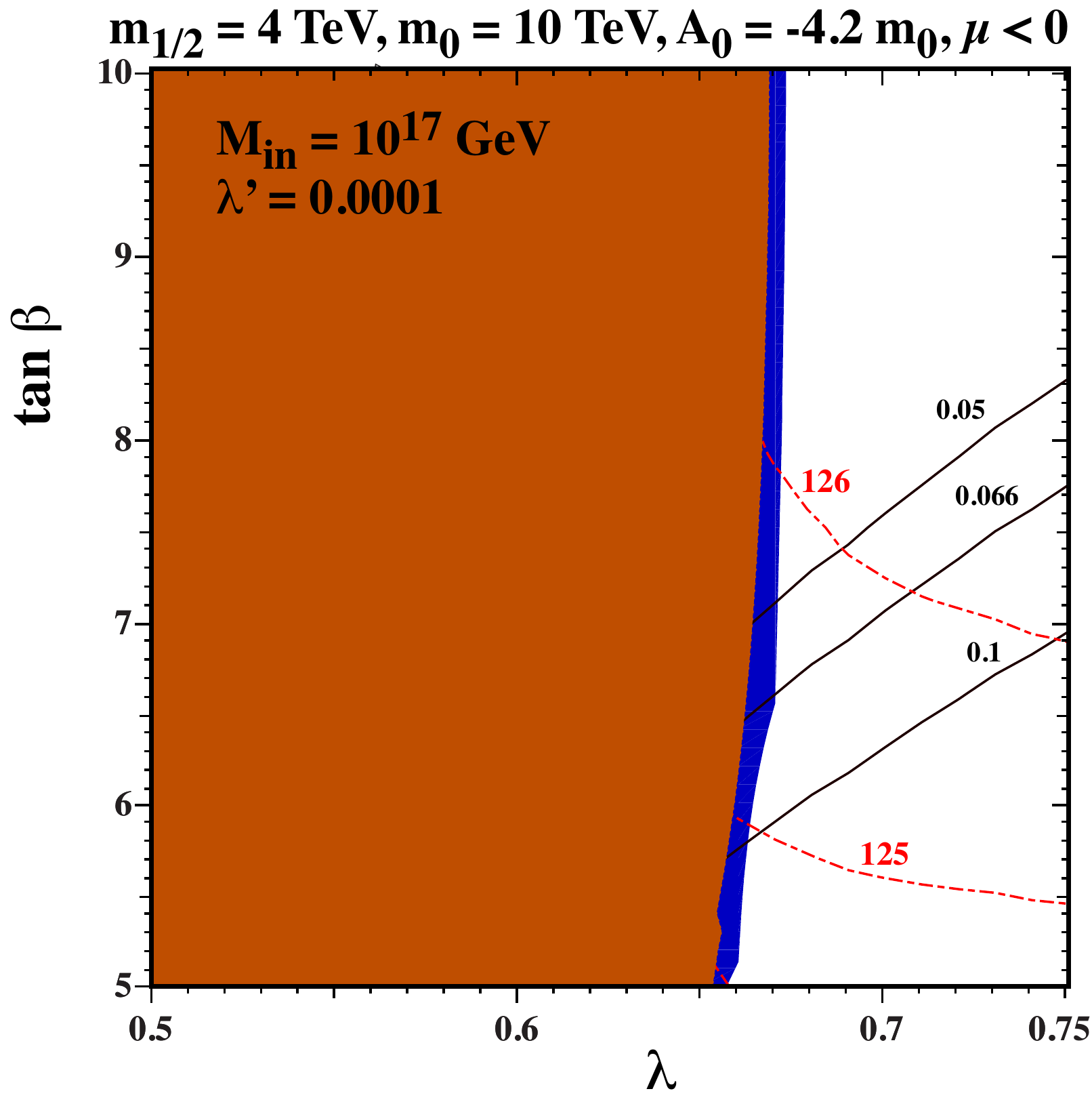}
\includegraphics[height=3.3in]{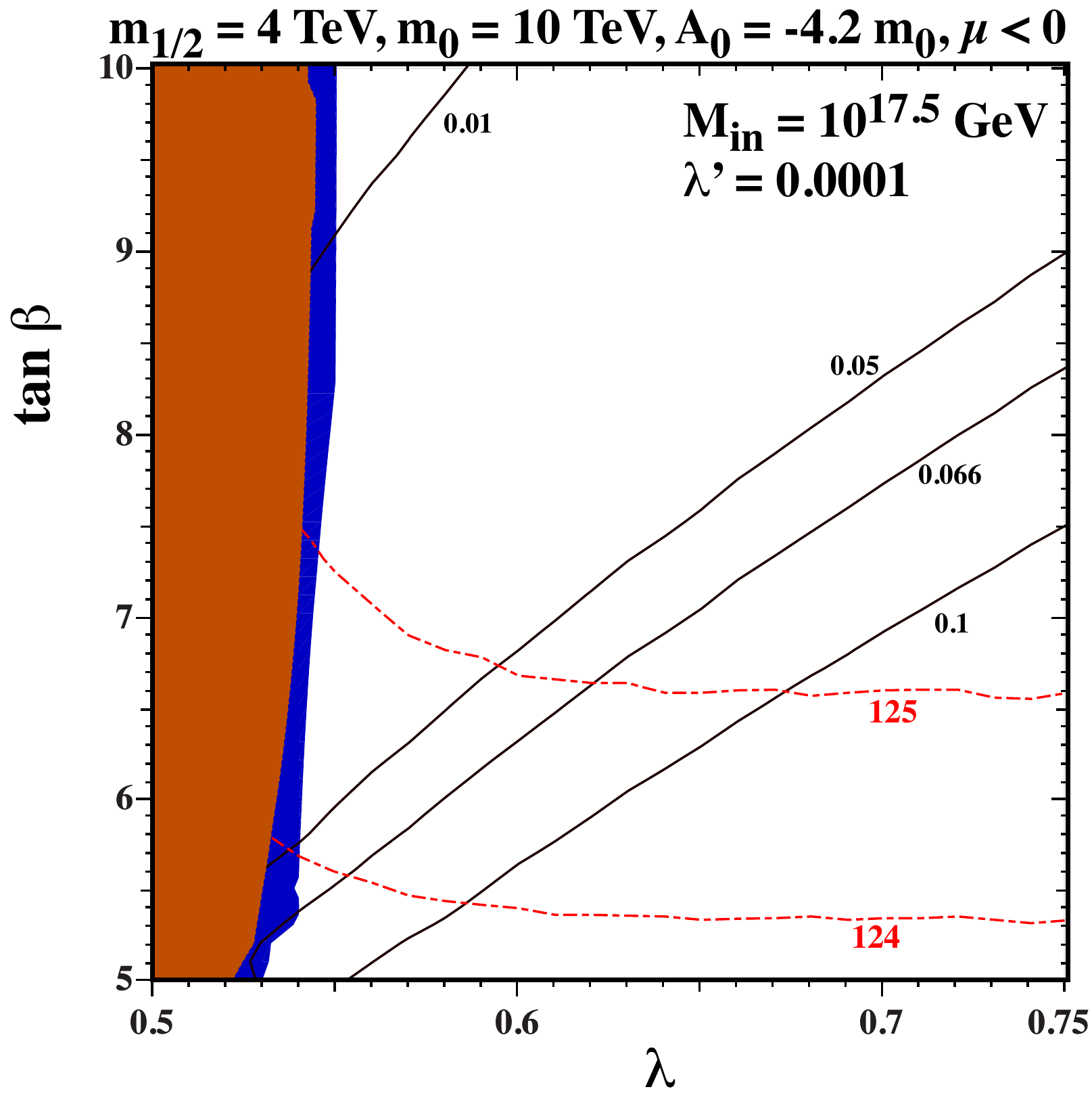}
\hfill
\end{minipage}
\begin{minipage}{8in}
\includegraphics[height=3.3in]{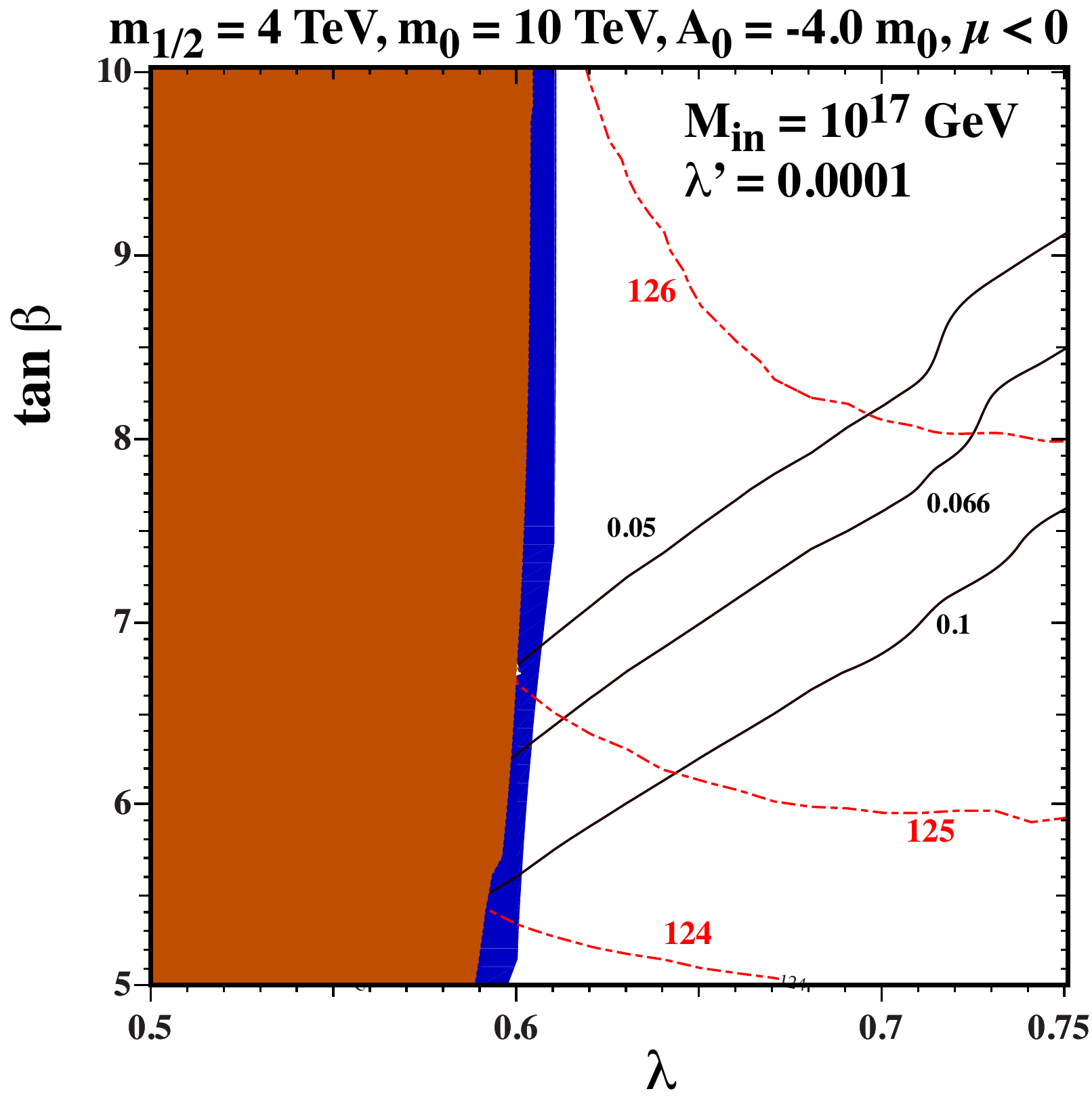}
\includegraphics[height=3.3in]{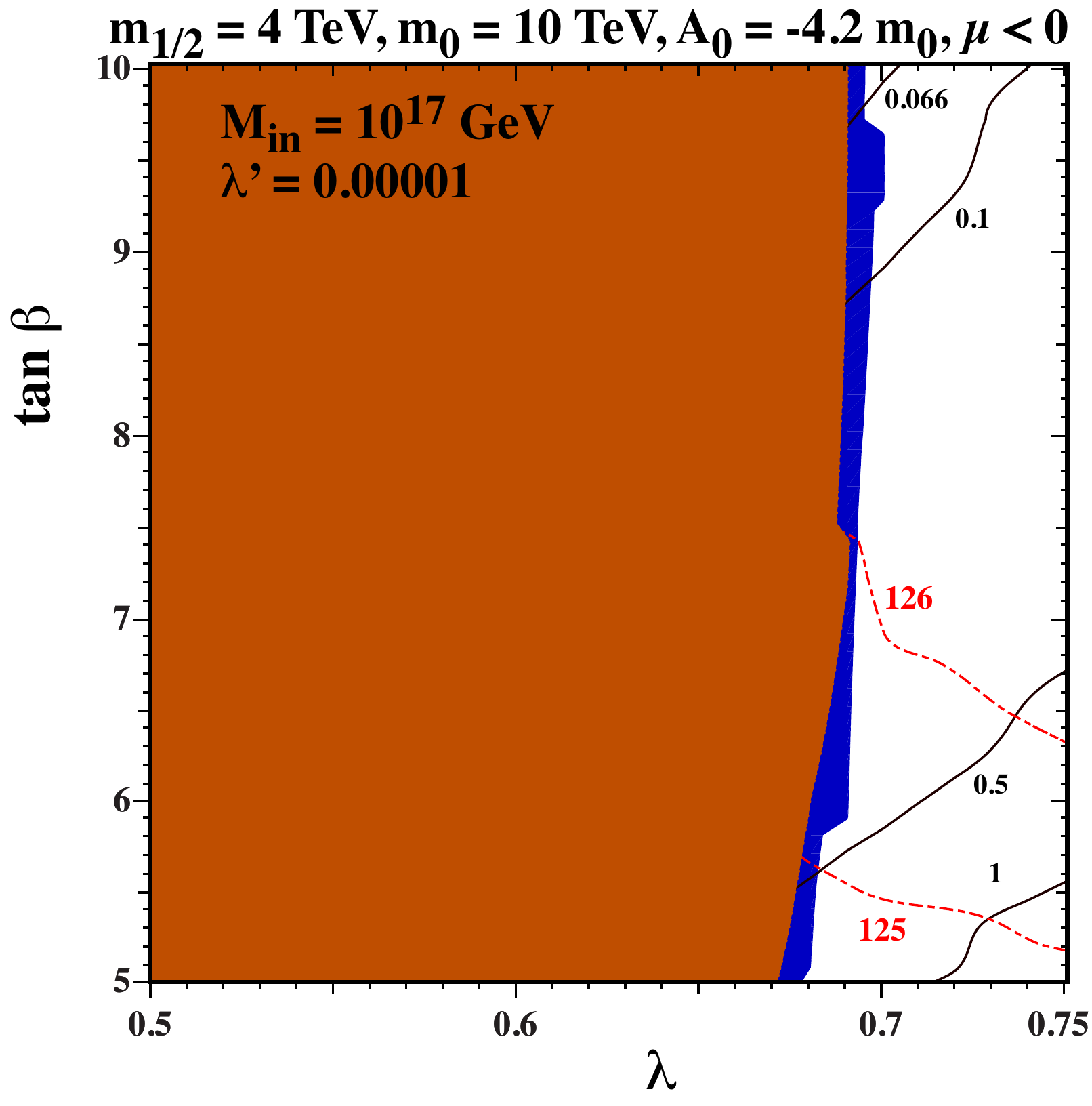}
\end{minipage}
\caption{
{\it
Super-GUT CMSSM $(\lambda,\tan \beta)$ planes with $m_{1/2} = 4$~TeV,
$m_0 = 10$~TeV, $\mu < 0$ and various values of $(M_{in}, A_0/m_0, \lambda^\prime)$
$ = (10^{17}$~GeV$, -4.2, 0.0001)$ (upper left), $ = (10^{17.5}$~GeV$, -4.2, 0.0001)$ (upper right),
$ = (10^{17}$~GeV$, -4.0, 0.0001)$ (lower left) and $ = (10^{17}$~GeV$, -4.2, 0.00001)$ (lower right).}
\label{fig:diskr2}}
\end{figure}

Finally, we discuss the dependence on $\lambda$ and $\lambda'$ by considering the
$(\lambda, \tan \beta)$ plots shown in Fig. \ref{fig:diskr2}, which are for $m_{1/2} = 4$~TeV,
$m_0 = 10$~TeV and $\mu < 0$, with different values of $(M_{in}, A_0/m_0, \lambda^\prime)$.
The upper left panel is with the values $(10^{17}$~GeV$, -4.2, 0.0001)$, which serve as
references. We see that the dark matter strip is adjacent to the brown stop LSP region
at $\lambda \simeq 0.67$, growing only slightly with $\tan \beta$ in the range displayed.
Along this strip, the proton lifetime constraints is respected for $\tan \beta \lesssim 6.5$,
where $m_h \sim 125$~GeV according to {\tt FeynHiggs}.
Here, one sees very clearly the dependences of $m_h$ and $\tau_p$ on $\tan \beta$.

In the upper right panel of Fig. \ref{fig:diskr2},
$M_{in}$ is increased to $(10^{17.5})$~GeV, and we see that the dark matter-compatible
value of $\lambda$ decreases to $\sim 0.55$ and proton stability then enforces $\tan \beta \lesssim 5.2$,
with $m_h$ about a GeV smaller than before, but still compatible with the LHC measurement
when the {\tt FeynHiggs} uncertainties are taken into account.
Had we decreased $M_{in}$ to $10^{16.5}$ GeV, the coannihilation strip would
have moved to $\lambda \approx 0.90$,
and the proton stability constraint would have required $\tan \beta \lesssim 8.3$. At the limit, $m_h \simeq 127$~GeV
and is lower at lower $\tan \beta$.

In the lower left panel of Fig. \ref{fig:diskr2},
$- A_0/m_0$ is decreased to 4.0, with $M_{in}$ and $\lambda^\prime$ taking their
reference values. In this case, the dark matter constraint requires $\lambda \sim 0.6$ and
proton stability then imposes $\tan \beta \lesssim 5.5$, again compatible with $m_h$.
Increasing $- A_0/m_0$ to 4.4 would move the coannihilation strip to $\lambda \simeq 0.72$, and
the limit on $\tan \beta$ would become $\tan \beta \lesssim 6.6$ with $m_h$ close to 126 GeV.

Finally, we see in the lower right panel of Fig. \ref{fig:diskr2} that for $\lambda^\prime = 0.00001$ and the reference
values of $M_{in}$ and $A_0/m_0$ the dark matter density requires $\lambda \simeq 0.68$
and proton stability then allows $\tan \beta \lesssim 9.8$. Most of this part of the strip
is also compatible with $m_h$, given the uncertainty in the {\tt FeynHiggs} calculation.
A larger value of $\lambda^\prime = 0.001$ would require $\tan \beta \lesssim 3.6$, but for this value of
$\tan \beta$ the Higgs mass would be unacceptably small, around 120.4 GeV.

\section{Discussion}
\label{sec:discussion}

It is frequently stated that the minimal SU(5) GUT model is excluded by
the experimental lower limit on the proton lifetime. Taking into account
the cosmological constraint on the cold dark matter density, the LHC
measurement of $m_h$ and the unknown GUT-scale phases appearing in the
SU(5) GUT model, we have shown in this paper that this model is quite
compatible with the proton stability constraint. 

We remind the reader that the amplitudes for the (normally) dominant
$p\to K^+ \overline{\nu}$ decay modes depend on two GUT-scale phases
that are beyond the CKM framework, and are not constrained by low-energy
physics. As we have discussed in detail, their effects on the $p\to K^+
\overline{\nu}_\tau$ decay amplitude are different from those on the
$p\to K^+ \overline{\nu}_{e, \mu}$ decay amplitudes. We take these
effects into account, and also consider their effects on the (normally)
subdominant $p\to \pi^+ \overline{\nu}$ and $n\to \pi^0 \overline{\nu}$
decays modes. In order to derive the most conservative bounds on the
model parameters, we choose the unknown GUT-scale phases so as to
maximize the $p\to K^+ \overline{\nu}$ lifetime. 

The compatibility of the supersymmetric GUT model with the proton
stability constraint is already visible in the CMSSM with universality
of the soft supersymmetry-breaking scalar masses imposed at an input
scale $M_{in} = M_{GUT}$ and $\tan \beta \sim 5$. This is visible in
Fig.~\ref{fig:CMSSM} along the upper parts of the focus-point strips in
the left panels (with $A_0 = 0$) and of the stop-coannihilation strips
in the right panels (with $A_0 = - 4.2 m_0$). According to the latest
version of {\tt FeynHiggs}, large portions of these strips are also
compatible with the experimental measurement of $m_h$. 

The super-GUT CMSSM with $M_{in} > M_{GUT}$ has more parameters, namely
the superpotential couplings $\lambda$ and $\lambda^\prime$ as well as
$M_{in}$. Correspondingly, the super-GUT CMSSM has greater scope for
compatibility with the proton stability and $m_h$ constraints. We had
previously noted \cite{emo} that, for $A_0 = 0$, the focus-point strip
move quickly to smaller $m_{1/2}$ and larger $m_0$ as $M_{in}$ is
increased. The stau LSP region also quickly recedes
\cite{superGUT,emo}. Here, we have added the matching condition for $B$,
previously neglected in other analyses. This led us to concentrate on
relatively large values of $|A_0/m_0|$. We have given some illustrative
examples of suitable parameter choices in Figs.~\ref{fig:diskr1},
\ref{fig:superG} and \ref{fig:diskr2}. Typical value of the model
parameters are $M_{in} = 10^{17}$~GeV, $m_{1/2} = 4$~TeV, $m_0 =
10$~TeV, $A_0/m_0 \sim - 4$, $\tan \beta \sim 5$, $\lambda \sim 0.6$ and
$\lambda^\prime \lesssim 0.0001$. 

To evade the proton decay constraints, squarks are required to be as
heavy as $\gtrsim 10$~TeV, which are hard to probe even at the 100~TeV
collider; see~\cite{EZ}, however. On the other hand, the gluino mass can be $\lesssim 10$~TeV,
which can be probed at the 100~TeV collider \cite{Golling:2016gvc}. 
Such heavy sparticle masses require fine-tuning at the
electroweak scale \cite{eenz}; at the expense of this, the simple models discussed
in this paper, the minimal SU(5) GUT with (super-GUT) CMSSM, are found
to be able to meet all the phenomenological requirements. Of course, by
extending the models and/or introducing more complicated mechanisms, we
may find a less fine-tuned sparticle spectrum with which the problems in
the minimal SU(5), such as the doublet-triplet splitting and the
dimension-five proton decay problems, can be evaded---this is beyond the
scope of the present work.

In view of the sensitivity of the proton lifetime to the unknown GUT-scale phases, it would
interesting to derive model predictions for them---another objective for theories of quark
and lepton mixing to bear in mind. Even more interesting would be to devise ways to
measure these phases experimentally. In principle, one way to do this would be to
measure the ratios of $p\to K^+ \overline{\nu}$, $p\to \pi^+ \overline{\nu}$ and
$n\to \pi^0 \overline{\nu}$ decay modes, as illustrated in Fig.~\ref{fig:nucleon}.

This may seem like a distant prospect, but let us remember that the
Hyper-Kamiokande project, in particular, has an estimated 90\% CL sensitivity to
$p\to K^+ \overline{\nu}$ at the level of $2.5 \times 10^{34}$~yrs~\cite{HyperK}. This covers the
range allowed in Fig.~\ref{fig:phasedep} for the reference point indicated by a star $(\bigstar)$
in Fig.~\ref{fig:diskr1}, and illustrates the capability of Hyper-Kamiokande to
probe the GUT-scale physics of proton decay. Let us be optimistic!

\section*{Acknowledgements}

The work of J.E. was supported in part by the UK STFC via the research grant ST/J002798/1.
The work of J.L.E., N.N. and K.A.O.
was supported in part by DOE grant DE-SC0011842 at the University of Minnesota.

\section*{Appendix}
\appendix

\label{sec:protondecaycalc}
\renewcommand{\theequation}{A.\arabic{equation}}
\setcounter{equation}{0}

In this Appendix we review briefly the calculation of nucleon decay
rates in the minimal supersymmetric SU(5) GUT. For more details, see
Refs.~\cite{Hisano:2013exa, Nagata:2013sba, evno, Ellis:2015rya}.

As mentioned in the text, in the minimal supersymmetric SU(5) GUT model, the dominant
contribution to proton decay is induced by the exchange of the
color-triplet Higgs multiplets through the Yukawa interactions. We
parametrize the SU(5) Yukawa couplings as follows:
\begin{equation}
 \left(h_{\bf 10}\right)_{ij} = e^{i\phi_i} \delta_{ij} h_{{\bf 10},
  i}~,~~~~~~
\left(h_{\overline{\bf 5}}\right)_{ij} = V_{ij}^* h_{\overline{\bf 5},
j} ~.
\end{equation}
In this basis, the MSSM matter superfields are embedded as
$\Psi_i \in \{Q_i, e^{-i\phi_i}\overline{u}_i, V_{ij}\overline{e}_j\}$
and $\Phi_i \in \{\overline{d}_i, L_i\}$. Upon integrating out the
color-triplet Higgs multiplets, we obtain the dimension-five
effective operators in Eq.~\eqref{eq:efflaggut} with the Wilson
coefficients in Eq.~\eqref{eq:wilson5}. These coefficients are then
evolved down to the SUSY scale $M_{\rm SUSY}$ according to one-loop
RGEs, which are presented in Ref.~\cite{Ellis:2015rya}.

At $M_{\rm SUSY}$, sfermions
in the dimension-five operators are integrated out via the wino- or
Higgsino-exchange one-loop diagrams. This gives rise to dimension-six
baryon-number-violating operators. Keeping only the dominant
contributions, we have
\begin{align}
 {\cal L}^{\text{eff}}_6&=C^{\widetilde{H}}_i {\cal O}_{1i33}
+ C^{\widetilde{W}}_{jk}\widetilde{\cal O}_{1jjk}
+ C^{\widetilde{W}}_{jk}\widetilde{\cal O}_{j1jk}
+ \overline{C}^{\widetilde{W}}_{jk}\widetilde{\cal O}_{jj1k}
~,
\label{eq:efflagdim6}
\end{align}
with
\begin{align}
 {\cal O}_{ijkl} &\equiv \epsilon_{abc}(u^a_{Ri}d^b_{Rj})
(Q_{Lk}^c \cdot L_{Ll}) ~, \nonumber \\
 \widetilde{\cal O}_{ijkl} &\equiv \epsilon_{abc} \epsilon^{\alpha\beta}
\epsilon^{\gamma\delta} (Q^a_{Li\alpha}Q^b_{Lj\gamma})
(Q_{Lk\delta}^c L_{Ll\beta}) ~,
\end{align}
corresponding to the $O^{(1)}$ and $\widetilde{O}^{(4)}$ in
Ref.~\cite{Abbott:1980zj}, respectively. Here, $i =1,2$, $j=2,3$, and
$k=1,2,3$. The coefficients in Eq.~\eqref{eq:efflagdim6} are given by
\begin{align}
 C_i^{\widetilde{H}}
(M_{\text{SUSY}})&=\frac{f_tf_\tau}{(4\pi)^2}C^{*331i}_{5R}(M_{\text{SUSY}})
F(\mu, m_{\widetilde{t}_R}^2,m_{\tau_R}^2)~, \nonumber \\
 C^{\widetilde{W}}_{jk}(M_{\text{SUSY}}) &=
\frac{\alpha_2}{4\pi}C^{jj1k}_{5L}(M_{\text{SUSY}})
[F(M_2, m_{\widetilde{Q}_1}^2,
 m_{\widetilde{Q}_j}^2) +F(M_2, m_{\widetilde{Q}_j}^2,
 m_{\widetilde{L}_k}^2)] ~, \nonumber \\
 \overline{C}^{\widetilde{W}}_{jk}(M_{\text{SUSY}}) &=
-\frac{3}{2}
\frac{\alpha_2}{4\pi}C^{jj1k}_{5L}(M_{\text{SUSY}})
[F(M_2, m_{\widetilde{Q}_j}^2,
 m_{\widetilde{Q}_j}^2) +F(M_2, m_{\widetilde{Q}_1}^2,
 m_{\widetilde{L}_k}^2)] ~,
\label{eq:wilsusymatch}
\end{align}
where $m_{\widetilde{Q}_j}$ and $m_{\widetilde{L}_k}$
are the left-handed squark and left-handed lepton masses, respectively,
and~\footnote{Notice that for $M \ll m_1 \simeq m_2 \simeq M_{\rm SUSY}$,
$F(M, m_1^2, m_2^2) \simeq M/M_{\rm SUSY}^2$, while for $M  \simeq m_1
\simeq m_2 \simeq M_{\rm SUSY}$, $F(M, m_1^2, m_2^2) \simeq 1/(2M_{\rm
SUSY})$.}
 \begin{align}
F(M, m_1^2, m_2^2) &\equiv
\frac{M}{m_1^2-m_2^2}
\biggl[
\frac{m_1^2}{m_1^2-M^2}\ln \biggl(\frac{m_1^2}{M^2}\biggr)
-\frac{m_2^2}{m_2^2-M^2}\ln \biggl(\frac{m_2^2}{M^2}\biggr)
\biggr]~.
\label{eq:funceq}
\end{align}
Note that the wino and Higgsino contributions are proportional to
$C_{5L}$ and $C_{5R}$, respectively. The coefficients in
Eq.~\eqref{eq:wilsusymatch} are then run down to the electroweak scale
by using one-loop RGEs \cite{Ellis:2015rya, Alonso:2014zka}.

We consider in this paper the $p\to K^+ \overline{\nu}$, $p\to \pi^+
\overline{\nu}$ and $n\to \pi^0 \overline{\nu}$ channels. Other nucleon
decay modes are less important, or their experimental limits are less
constraining. The effective interactions for the $p\to K^+ \overline{\nu}$ is
given by
\begin{align}
 {\cal L}(p\to K^+\bar{\nu}_i^{})
&=C_{RL}(usd\nu_i)\bigl[\epsilon_{abc}(u_R^as_R^b)(d_L^c\nu_i^{})\bigr]
+C_{RL}(uds\nu_i)\bigl[\epsilon_{abc}(u_R^ad_R^b)(s_L^c\nu_i^{})\bigr]
\nonumber \\
&+C_{LL}(usd\nu_i)\bigl[\epsilon_{abc}(u_L^as_L^b)(d_L^c\nu_i^{})\bigr]
+C_{LL}(uds\nu_i)\bigl[\epsilon_{abc}(u_L^ad_L^b)(s_L^c\nu_i^{})\bigr]
~,
\end{align}
while the $p\to \pi^+ \overline{\nu}$ and $n\to \pi^0 \overline{\nu}$
channels are induced by
\begin{align}
  {\cal L}(p\to \pi^+ \bar{\nu}_i)
= C_{RL}(udd\nu_i)\bigl[\epsilon_{abc}(u_R^ad_R^b)(d_L^c\nu_{Li}^{})\bigr]
+C_{LL}(udd\nu_i)\bigl[\epsilon_{abc}
(u_L^ad_L^b)(d_L^c\nu_{Li}^{})\bigr]
~.
\label{eq:leffppinu}
\end{align}
These Wilson coefficients are evaluated at the weak scale as follows:
\begin{align}
 C_{RL}(usd\nu_\tau)&=-V_{td}C^{\widetilde{H}}_{2}(m_Z)~,\nonumber \\
 C_{RL}(uds\nu_\tau)&=-V_{ts}C^{\widetilde{H}}_{1}(m_Z)~,\nonumber \\
 C_{RL}(udd\nu_\tau)&=-V_{td}C^{\widetilde{H}}_1 (m_Z)~,\nonumber \\
 C_{LL}(udd\nu_k)&=\sum_{j=2,3} V_{j1} V_{j1} C_{jk}^{\widetilde{W}}
(m_Z) ~, \nonumber \\
 C_{LL}(usd\nu_k)&=\sum_{j=2,3}V_{j1}V_{j2}
C^{\widetilde{W}}_{jk}(m_Z)~,\nonumber \\
 C_{LL}(uds\nu_k)&=\sum_{j=2,3}V_{j1}V_{j2}
C^{\widetilde{W}}_{jk}(m_Z) ~.
\end{align}
We note that the $C_{RL}$ and $C_{LL}$ coefficients are induced by the
Higgsino and wino contributions, respectively.

\begin{table}[t]
\centering
\caption{\it Hadron matrix elements for nucleon decay. See
 Ref.~\cite{Aoki:2013yxa} for computations of these values, including error estimates.
}
\vspace{3mm}
\label{tab:matrixelements}
\begin{tabular}{lc|lc}
\hline \hline
Matrix element&Value (GeV$^2$)& Matrix element&Value (GeV$^2$) \\[2pt]
\hline
$\langle K^+\vert (us)_L ^{}d_L^{}\vert p\rangle$& 0.036
&$\langle \pi^+\vert (ud)_R^{}d_L^{}\vert p\rangle$& $-0.146$
\\
$\langle K^+\vert (ud)_L ^{}s_L^{}\vert p\rangle$& 0.111
&$\langle \pi^+\vert (ud)_L^{}d_L^{}\vert p\rangle$& 0.188
\\
$\langle K^+\vert (us)_R ^{}d_L^{}\vert p\rangle$& $-0.054$
&$\langle \pi^0\vert (ud)_R^{}d_L^{}\vert n\rangle$& $-0.103$
\\
$\langle K^+\vert (ud)_R ^{}s_L^{}\vert p\rangle$& $-0.093$
&$\langle \pi^0\vert (ud)_L^{}d_L^{}\vert n\rangle$& 0.133
\\
\hline
\hline
\end{tabular}
\end{table}

Using the two-loop RGEs given in Ref.~\cite{Nihei:1994tx}, we evolve
these coefficients down to the hadronic scale $\mu_{\text{had}}=
2$~GeV, where the matrix elements of the effective operators are
evaluated. Values of the relevant hadron matrix elements are summarized in
Table~\ref{tab:matrixelements}, as computed using QCD
lattice simulations in Ref.~\cite{Aoki:2013yxa}. The decay width of each
decay channel is then given by
\begin{align}
  \Gamma(p\to K^+\bar{\nu}_i)
&=\frac{m_p}{32\pi}\biggl(1-\frac{m_K^2}{m_p^2}\biggr)^2
\vert {\cal A}(p\to K^+\bar{\nu}_i)\vert^2~, \\
 \Gamma (p\to  \pi^+ \bar{\nu}_i) &=
\frac{m_p}{32\pi}\biggl(1-\frac{m_\pi^2}{m_p^2}\biggr)^2
\vert {\cal A}(p\to \pi^+ \bar{\nu}_i) \vert^2~, \\
 \Gamma (n\to  \pi^0 \bar{\nu}_i)&=
\frac{m_n}{32\pi}\biggl(1-\frac{m_\pi^2}{m_n^2}\biggr)^2
\vert {\cal A}(n\to \pi^0 \bar{\nu}_i) \vert^2~,
\end{align}
where $m_p$, $m_n$, $m_K$, and $m_\pi$ are the masses of the proton,
neutron, kaon, and pion, respectively, and
\begin{align}
 {\cal A}(p\to K^+\bar{\nu}_i)&=
C_{RL}(usd\nu_i)\langle K^+\vert (us)_Rd_L\vert p\rangle
+
C_{RL}(uds\nu_i)\langle K^+\vert (ud)_Rs_L\vert p\rangle
\nonumber \\
&+
C_{LL}(usd\nu_i)\langle K^+\vert (us)_Ld_L\vert p\rangle
+C_{LL}(uds\nu_i)\langle K^+\vert (ud)_Ls_L\vert p\rangle
~, \nonumber\\[2pt]
 {\cal A} (p\to \pi^+ \bar{\nu}_i)&=
C_{RL}(udd\nu_i)\langle \pi^+\vert (ud)_Rd_L\vert p\rangle
+C_{LL}(udd\nu_i)\langle \pi^+\vert (ud)_Ld_L\vert p\rangle ~, \nonumber\\[2pt]
 {\cal A} (n\to \pi^0 \bar{\nu}_i)&=
C_{RL}(udd\nu_i)\langle \pi^0\vert (ud)_Rd_L\vert n\rangle
+C_{LL}(udd\nu_i)\langle \pi^0\vert (ud)_Ld_L\vert n\rangle~.
\label{eq:amplitudes}
\end{align}
We note that the $C_{RL}$ coefficients are non-vanishing only for
$i=\tau$. Thus, the decay channels that contain $\overline{\nu}_e$ or
$\overline{\nu}_\mu$ are induced by wino exchange only.



\end{document}